\documentclass[fleqn,usenatbib]{mnras}

\usepackage{newtxtext,newtxmath}

\usepackage[T1]{fontenc}

\DeclareRobustCommand{\VAN}[3]{#2}
\let\VANthebibliography\thebibliography
\def\thebibliography{\DeclareRobustCommand{\VAN}[3]{##3}\VANthebibliography}

\newcommand{\hi}{\textrm{H\textsc{i}}}
\newcommand{\secref}[1]{\hyperref[#1]{Section~\ref*{#1}}}
\newcommand{\appref}[1]{\hyperref[#1]{Appendix~\ref*{#1}}}

\usepackage{graphicx}	
\usepackage{amsmath}	
\usepackage{ctable} 
\usepackage{enumitem}

\title[Regridding LIM for FFTs: mitigating aliasing
]
{Accurate Fourier-space statistics for line intensity mapping: Cartesian grid sampling without aliased power}

\author[S. Cunnington \& L. Wolz]{Steven Cunnington$^1$\thanks{E-mail: \href{mailto:steven.cunnington@manchester.ac.uk}{steven.cunnington@manchester.ac.uk}} \& Laura Wolz$^1$
\\
$^1$Jodrell Bank Centre for Astrophysics, Department of Physics \& Astronomy, The University of Manchester, Manchester M13 9PL, UK}

\date{Accepted XXX. Received YYY; in original form ZZZ}

\pubyear{2023}

\begin{document}
\label{firstpage}
\pagerange{\pageref{firstpage}--\pageref{lastpage}}
\maketitle

\begin{abstract}
Estimators for $n$-point clustering statistics in Fourier-space demand that modern surveys of large-scale structure be transformed to Cartesian coordinates to perform Fast Fourier Transforms (FFTs). In this work, we explore this transformation in the context of pixelised line intensity maps (LIM), highlighting potential biasing effects on power spectrum measurements. Current analyses often avoid a complete resampling of the data by approximating survey geometry as rectangular in Cartesian space, an increasingly inaccurate assumption for modern wide-sky surveys. Our simulations of a $20\,{\times}\,20\,\text{deg}^2$ 21cm LIM survey at $0.34\,{<}\,z\,{<}\,0.54$ show this assumption biases power spectrum measurements by ${>}\,20\%$ across all scales. We therefore present a more robust framework for regridding the voxel intensities onto a 3D FFT field by coordinate transforming large numbers of Monte-Carlo sampling particles. Whilst this unbiases power spectrum measurements on large scales, smaller-scale discrepancies remain, caused by structure smoothing and aliasing from separations unresolved by the grid. To correct these effects, we introduce modelling techniques, higher-order particle assignments, and interlaced FFT grids to suppress the aliased power. Using a Piecewise Cubic Spline (PCS) particle assignment and an interlaced FFT field, we achieve sub-percent accuracy up to 80\% of the Nyquist frequency for our 21cm LIM simulations. We find a more subtle hierarchical improvement in results for higher-order assignment schemes, relative to the gains made for galaxy surveys, which we attribute to the extra complexity in LIM from additional discretising steps. \texttt{Python} code accompanying this paper is available at \href{https://github.com/stevecunnington/gridimp}{\texttt{github.com/stevecunnington/gridimp}}.
\end{abstract}




\section{Introduction}

Surveying the integrated emission of spectral lines from galaxies is an emerging strategy for probing large-scale cosmic structure \citep{Bharadwaj:2000av,Battye:2004re,Visbal:2010rz,Visbal:2011ee, Fonseca:2016qqw, Kovetz:2017agg, Bernal:2022jap}. \textit{Line intensity mapping} (LIM) can efficiently chart density structure spanning large scales and redshift ranges. Examples of common spectral lines chosen for LIM are; 21cm from neutral hydrogen (\hi), rotational carbon-monoxide
(CO) transitions, fine-structure lines
such as CII, and shorter wavelength spectral lines such as H$\alpha$ and Ly$\alpha$ in the optical and ultraviolet. New instruments are returning pioneering data that demonstrate the potential LIM offers \cite{Cleary:2021dsp,Wang:2020lkn,HERA:2021bsv,Cunnington:2022uzo,CHIME:2022kvg,CONCERTO,Paul:2023yrr,Li:2023zer}.

Parameter inference from LIM will require clustering statistics such as $n$-point correlation functions to which models can be fitted. Optimal estimators often transform observations into Fourier space which allows faster computation. Furthermore, in the linear regime, Fourier modes will evolve independently, thus simplifying modelling since perturbations of different length scales are less correlated in Fourier space \citep{PeeblesBook}.

Whilst it is theoretically possible to transform all sky voxel coordinates into Cartesian coordinates and perform a direct summation to obtain the Fourier modes of the fluctuation field, the number of voxels within forthcoming intensity maps will render the direct summation method computationally unfeasible. For this reason, discrete Fast Fourier Transform (FFT) algorithms \cite{Cooley1965AnAF} will be needed which require sampling the LIM data onto a regular grid with Cartesian coordinates. A similar process happens with galaxy surveys whose catalogues of galaxy positions in 3D are too numerous for direct summation and thus must also be assigned to a grid to run an FFT. 

There has yet to be a dedicated study into the regridding of LIM data with sky coordinates (R.A., Dec. $\nu$) onto a regular 3D Cartesian grid with lengths $l_\text{x}, l_\text{y}, l_\text{z}$ in $h^{-1}$Mpc. Previous studies have approximated the LIM sky footprint to be rectangular, thus assuming equal voxel size everywhere and assigning Cartesian lengths to the map edges \citep[e.g.][]{Dillon:2012wx,Wolz:2015lwa,COMAP:2021sqw}. This allows an FFT on the map directly without the need for resampling the voxel intensities onto a new grid. This can be sufficient for small sky surveys, but as future LIM observations begin to span wider angles, this approximation becomes increasingly inaccurate. One way to resample the data is with a Monte-Carlo integration technique whereby sampling particles are distributed across the map and assigned the intensity values of the voxels they fall within. The position of these particles are then transformed to Cartesian coordinates and distributed into cells on a grid which can be FFTed. The Monte-Carlo sampling technique has mainly been used in simulations for doing the reverse; sampling a Cartesian cosmological cube into spherical sky voxels \citep[e.g.][]{Alonso:2014sna}. In \citet{Blake:2019ddd}, a Monte-Carlo sampling was utilised in simulations for validating the modelling of observational effects for \hi\ intensity mapping, but a dedicated study into its performance as a regridding tool is still warranted.

A concern for any regridding technique is how it will distort the true field properties from the raw observations. For example, aliased contributions can be introduced by perturbations across small separations no longer resolved by the finite size of the new grid cells. This has been previously studied for radio interferometry where converting visibilities into image space requires sampling the visibility values onto a grid so an FFT can be performed to produce the image. Early experiments used straightforward nearest neighbour schemes for assigning the visibilities into cells on a Cartesian grid \citep{HoggPaper}, but this was shown to produce unwanted aliasing. Convolutions with more sophisticated window functions have since been explored \citep{CASA,2020MNRAS.491.1146Y,2018A&A...616A..27V,9897317,Barry:2022mnm}, where often the instrumental response of the interferometer needs to be incorporated into the convolution kernel. 

For auto-correlator (e.g. \textit{single-dish} or single-receiver) experiments such as SPHEREx, FAST, MeerKAT, SKAO, COMAP \cite{SPHEREx:2014bgr,FAST,MeerKLASS:2017vgf,Bacon:2018dui,Cleary:2021dsp}, LIM observations are quite different. Whilst interferometers make angular measurements directly in Fourier space, auto-correlator observations provide time-ordered data which is made into sky maps in \textit{real}- (or configuration)-space, represented by 3D arrays of voxels, each with sky angular coordinates and frequencies corresponding to the redshifted spectral line. Resampling these voxel positions which carry intensities is akin to resolved galaxy surveys which are also required to assign weighted point-like galaxies with 3D sky coordinates onto a regularly spaced Cartesian grid for an FFT.

The effects of discretised gridding on the power spectra of large galaxy catalogues have been well studied \citep{Jing:2004fq,Cui:2008fi,2009RAA.....9..227Y,Colombi:2008dw,Jasche:2009kk,Sefusatti:2015aex}. Aliasing in the power spectrum arises from a periodic summation of contributions at multiples of the sampling frequency of the grid. Techniques can be used such as higher-order interpolations beyond a zero-order Nearest Grid Point (NGP) mass assignment \citep{HockneyEastwood}. These are effectively low-pass filters that suppress the contributions from the unresolved small-scale modes but have to be increasingly non-local to entirely remove aliasing. Any mass assignment choice requires some correction to the power spectrum since the measured power spectrum from an FFT grid will not be equivalent to the true power spectrum of the raw field, but instead, one that is convolved with a window function $W^2_{\rm mas}(\boldsymbol{k})$ whose shape is determined by the mass assignment scheme adopted \citep{Jing:2004fq}. Other techniques such as interlacing the FFT grids have also been shown to remove the odd summations of the sampling frequency multipoles and suppress aliasing \citep{Sefusatti:2015aex}. 

In this work, we use simulations to examine the performance of Monte-Carlo sampling techniques for regridding LIM data into Cartesian coordinates. We present a modelling pipeline that approximates many of the distorting effects from the mapping and then regridding of the LIM field. We also explore some of the techniques used by galaxy survey analyses, such as higher-order mass assignment and interlacing, to see if they can be applied to the Monte-Carlo sampling particles to mitigate the aliased power they introduce.  

The paper structure sees \secref{sec:RegridMonte} detail the simulations we adopt and the Monte-Carlo-style sampling process used to transform the LIM data onto the final Cartesian grid. We demonstrate the distortions to power spectrum measurements caused by this process if no corrections are applied.
\secref{sec:MitigatingAliasingContributions} focuses on improving the accuracy of the power spectrum measurements relative to a modelled expectation. This is done by extending the modelling to include various discretiastion effects as well as including higher-order interpolation schemes for the sampling particles and interlacing the FFT fields. We discuss and conclude the results in \secref{sec:Conclusion}.\newline
\\
\noindent We adopt some consistent language and notation conventions throughout the paper; referring to 3D discretised field bins as \textit{voxels} when in sky (RA., Dec., $\nu$) coordinates, and \textit{cells} when in Cartesian (x, y, z) coordinates. We also sometimes distinguish between angular and radial binning of sky data with \textit{pixels} and frequency \textit{channels}, respectively. We use the Greek vector $\boldsymbol{\theta}$ to refer to an angular position in sky coordinates at a frequency $\nu$, and the Latin vector $\boldsymbol{x}$ for a position in Cartesian space. We refer to the input LIM fluctuation field (in Cartesian space) by $\delta_0(\boldsymbol{x})$, the array of \texttt{HEALPix} maps at each frequency with $m(\boldsymbol{\theta},\nu)$, and the final resampled Cartesian grid, on which the FFT is performed, as $\delta_\text{G}(\boldsymbol{x})$.

\section{LIM Regridding with Monte Carlo sampling }\label{sec:RegridMonte}

Normalized temperature fluctuations, $\delta_\text{LIM}$, from line emitters are a biased tracer of the underlying dark matter field $\delta_\text{m}$, in the post-epoch of reionisation era. This is simply expressed with
\begin{equation}
    \delta_\text{LIM}(\boldsymbol{x},z) = b(z)\delta_\text{m}(\boldsymbol{x},z)\,,
\end{equation}
where $b$ is the linear bias of the lines brightness temperature field. A LIM experiment will observe brightness temperature fluctuations given by
\begin{equation}
    \delta T_\text{LIM}(\boldsymbol{x},z) = T(\boldsymbol{x},z)-\overline{T}(z) = \delta_\text{LIM}(\boldsymbol{x},z) \overline{T}(z)\,,
\end{equation}
where $T$ are the absolute temperatures of the field, generally not accessible by a LIM experiment, and $\overline{T}$ is the mean brightness temperature. Surveys record the continuous temperature fluctuations, $\delta T_\text{LIM}$, binning time-ordered data into voxels on a sky map $m(\boldsymbol{\theta},\nu)$, using a scheme such as \texttt{HEALPix} \citep{Gorski:2004by} pixelisation for all frequencies within the survey's wavebands. A Fourier-space analysis of the temperature fluctuations, such as a power spectrum estimation, will be imprecise if done directly on the sky maps $m$ since its geometry is not a regular spaced grid in Cartesian space. A resampling process onto a Cartesian grid $\delta_\text{G}(\boldsymbol{x})$ is therefore required to coordinate transform the collection of voxels with angular pointings $\boldsymbol{\theta}$ (R.A. and Dec.) and frequency $\nu$ (MHz), related to redshift through $z\,{=}\,\nu_0/\nu\,{-}\,1$, where $\nu_0$ is the rest frame frequency of the target spectral line. The regridded data in Cartesian space can then be discrete Fourier-transformed following
\begin{equation}\label{eq:deltaG_k}
    \tilde{\delta}_\text{G}(\boldsymbol{k})=\sum_{\boldsymbol{x}} \delta_\text{G}(\boldsymbol{x}) w(\boldsymbol{x}) \exp (i \boldsymbol{k} \cdot \boldsymbol{x})\,,
\end{equation}
where $w$ are weights used to optimise the analysis. In this work, for simplicity, we use unity weighting but inverse thermal noise weighting is more applicable in real data (see \citet{Blake:2019ddd} for a derivation of optimal weighting for LIM). An estimator for any Fourier-space $n$-point clustering statistic can then be applied to $\tilde{\delta}_\text{G}$ such as a power spectrum estimation, given by
\begin{equation}
    \hat{P}_\text{G}(\boldsymbol{k})=\frac{V_{\text {cell }}}{\sum_{\boldsymbol{x}} w^2(\boldsymbol{x})}\left|\tilde{\delta}_\text{G}(\boldsymbol{k})\right|^2\,,
\end{equation}
where $V_\text{cell}$ is the volume of a single cell in the regridded field.

\begin{figure*}
    \centering
    \includegraphics[width=1\linewidth]{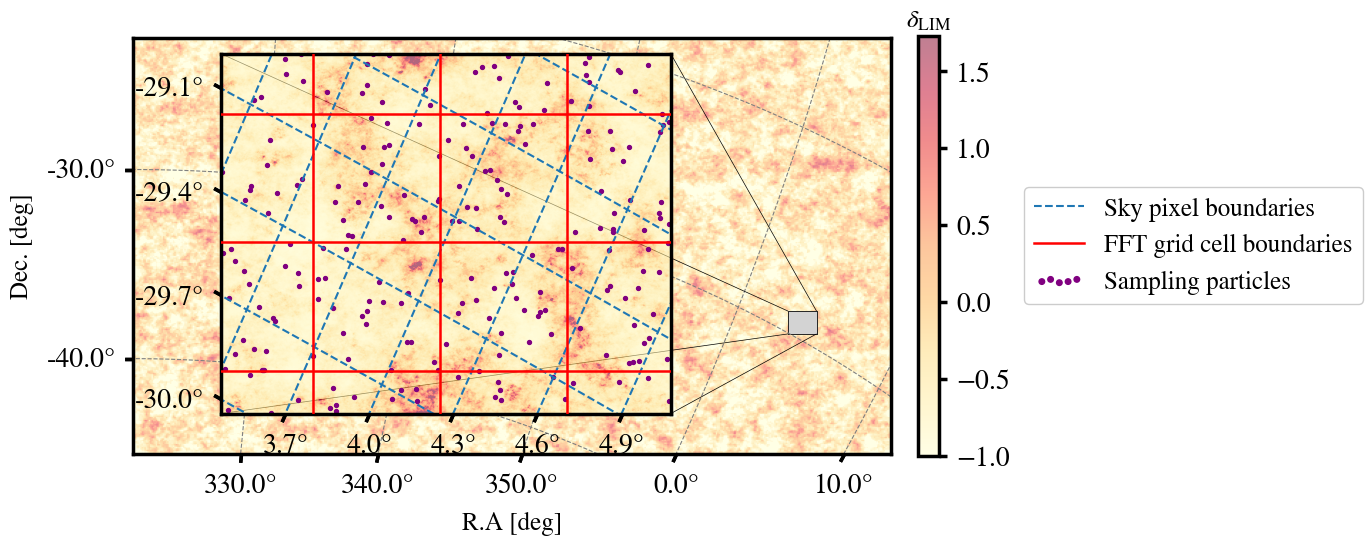}
    \caption{Toy illustration of LIM pixelisation and Cartesian gridding processes necessary for analysis in Fourier-space. The background image represents a continuous fluctuation field for the spectral line (field is mean-centred and mean-normalised). LIM surveys will discretise their calibrated observations of the continuous field into broad sky pixels (blue dashed boundaries in insert). This provides the 3D voxel array in sky coordinates (R.A., Dec, $\nu$). Using a Monte-Carlo method with sampling particles, the sky voxel intensities can be regridded onto a 3D Cartesian grid (red boundaries) in comoving space ($l_\text{x}$, $l_\text{y}$, $l_\text{z}$ in $h^{-1}\text{Mpc}$) which can be Fast Fourier transformed. The method we adopt (detailed in \secref{sec:MoneteCarlosummary}) creates \textit{pseudo}-random particles by generating $N_\text{p}$ per input sky voxel ($N_\text{p}\,{=}\,10$ for this example), with one particle at the voxel centre and the other $N_\text{p}-1$ distributed uniformly randomly within the voxel.}
    \label{fig:MKmap2grid}
\end{figure*}

In \autoref{fig:MKmap2grid} we sketch out the process of sky pixelisation in a real LIM data reduction pipeline, followed by a regridding into Cartesian cells (following a Monte-Carlo sampling technique which we detail shortly). This shows how the continuous emission of the spectral feature is discretised into sky pixels, which are represented by the blue-dashed lines. Depending on the instrument and map-making choices, these can be quite broad and the $0.3\,\text{deg}$ size pixels shown are representative of the MeerKAT 21cm IM pilot survey data \citep{Wang:2020lkn}. Once pixelised maps have been created for each frequency channel of the instrument, the stacked maps need to be resampled onto the Cartesian grid. Example grid boundaries are shown in \autoref{fig:MKmap2grid} by the red-solid lines, the size of which are free to be chosen, which we discuss shortly. The plot also shows example Monte-Carlo sampling particles. Each particle is assigned a sampling weight, equal to the intensity of its host sky map voxel. In this example, we use 10 for the number of sampling particles per map voxel ($N_\text{p}$). These particles can then be trivially transformed\footnote{For example \href{https://docs.astropy.org/en/stable/index.html}{\texttt{astropy}} \citep{Astropy} has routines for this conversion which we utilise in this work.} and their intensities populated onto the red grid in Cartesian space. A Cartesian grid cell value is determined by the average intensity from all sampling particles that fall within the cell (at least for the nearest grid point assignment scheme). \autoref{fig:MKmap2grid} only shows the 2D collapsed projection in angular space but the discretisation into voxels and resampling onto a Cartesian grid is 3-dimensional. 

\subsection{Monte-Carlo sampling summary}\label{sec:MoneteCarlosummary}

Here we detail the exact steps we follow for performing a resampling of the pixelised LIM onto a Cartesian grid. The scatter points in \autoref{fig:MKmap2grid} demonstrate the pseudo-random Monte Carlo sampling technique we adopt in this work whereby a constant number ($N_\text{p}$) of particles are assigned to each voxel in the sky map. The exact sampling process is summarised by the following steps:

\begin{itemize}
    \item $N_\text{p}$ particles per input sky voxel are generated with one at the voxel centre and the remaining $N_\text{p}\,{-}\,1$ particles randomly distributed with uniform probability density within the voxel.
    \item Each $i$th particle's mass is assigned the full amplitude $m(\theta_i)$ of the sky voxel it originated from (this is later normalised in the new Cartesian grid).
    \item The (R.A., Dec., $\nu$) sky coordinates of the particles are transformed into (x, y, z in $h^{-1}\text{Mpc}$) Cartesian coordinates.
    \item At this point, an interpolation scheme must be chosen which we explore in detail later. For now, we assume a nearest grid point assignment where the full mass of every $i$th particle is added to the Cartesian cell $\delta_\text{G}(x_i)$ it falls within.
    \item We normalise all Cartesian cells by dividing by the count of particles that fall within it. Therefore the amplitude of each Cartesian cell will be given by
    \begin{equation}
        \delta_\text{G}(\boldsymbol{x}) = \frac{\sum_{i\in\boldsymbol{x}} w_{\text{p},i}\,m(\theta_i)}{\sum_{i\in\boldsymbol{x}}w_{\text{p},i}}\,,
    \end{equation}
    where $w_\text{p}$ provides the option for weighting each sampling particle (in this work we assume $w_\text{p}\,{=}\,1$ for all particles\footnote{We briefly explored assigning a weight based on the position of the particle relative to the voxel centre, however, this showed no obvious improvement. Further investigation is warranted though to see if this, or other weighting schemes, can further optimise results.}). An alternative choice would be to normalise the masses (or intensities) assigned to each sampling particle i.e. assigning $m(\theta_i)/N_\text{p}$. However, we found more consistent performance with unnormalised particles and a Cartesian grid cell averaging. We provide an analytical sketch motivation for this choice in \appref{sec:ResampledNormalisation}.
\end{itemize}

This approach differs subtly from previous work that also utilised a Monte-Carlo-style integration for regridding. \citet{Blake:2019ddd} used a fully random process whereby a large amount of particles are randomly distributed across the survey sky map. We found our choice of using a constant number of sampling particles per input map voxel yielded more consistent results across different scenarios. This is likely because the intensities from the map are being divided by consistent quantities i.e. the number of sampling particles generated for each voxel is independent of $\boldsymbol{\theta}$, the position on the map. This is not necessarily the case in a fully randomised sampling process since an extra element of stochasticity is introduced. The sampling count in each voxel would also vary with $\boldsymbol{\theta}$ if we opted to generate a constant number of sampling particles per \textit{output} Cartesian cell, causing position-dependent averaging problems.

\begin{figure}
    \centering
    \includegraphics[width=1\linewidth]{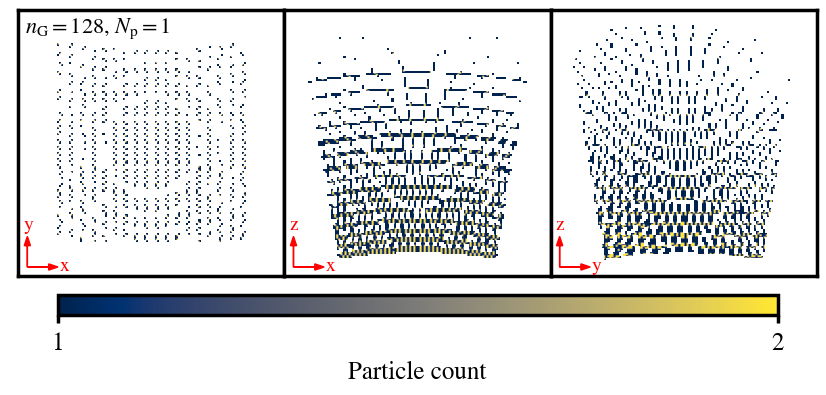}
    \includegraphics[width=1\linewidth]{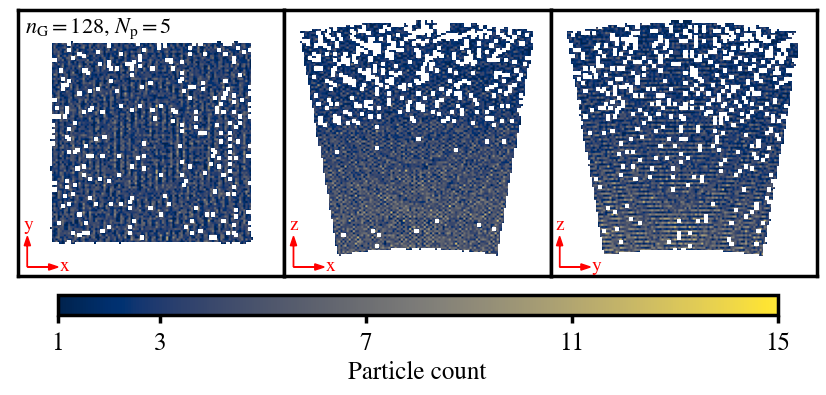}
    \includegraphics[width=1\linewidth]{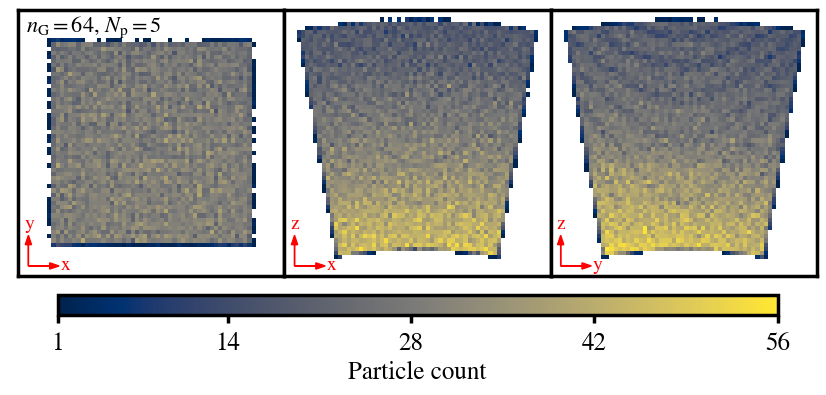}
    \caption{Sampling coverage on the final Cartesian grid for a toy $20\,{\times}\,20\,\text{deg}$ simulated 21cm LIM survey spanning $900\,{<}\,\nu\,{<}1038\,\text{MHz}$ ($0.37\,{<}\,z\,{<}\,0.58$). The colour scale show the counts of sampling particles falling in the grid cells. Every panel shows a slice halfway through the array along the z,y and x directions going left to right, as indicated by the red directional arrows. Top-left text in each row gives details for each scenario. $n_\text{G}$ is the number of cells on the side of the cubic Cartesian grid. $N_{\rm p}$ is the number of sampling particles per pixel of the input map being regridded.}
    \label{fig:MonteCarloDemo}
\end{figure}

The choice of cell size in Cartesian space (red grid in \autoref{fig:MKmap2grid}), is a balance between ensuring enough sampling particles fall within all cells across the grid but not making the cells too large and further limiting the Nyquist frequency in the analysis. One could set $N_\text{p}\,{=}\,1$ which is equivalent to treating the map voxel centres as a discrete set of points whose coordinates can be transformed to Cartesian space. However, if this is done on too fine a grid, then there will be empty cells within the footprint of the survey, which creates ringing artifacts and mode coupling in the Fourier transform which are difficult to model. \autoref{fig:MonteCarloDemo} illustrates the empty cell effect with a toy example; a $20\,{\times}\,20\,\text{deg}$ simulated survey spanning $900\,{<}\,\nu\,{<}1038\,\text{MHz}$ (details of our main simulations are explained in the following sub-section). Choosing to simply transform the sky map voxels and bin them into the nearest FFT cells on a $n_\text{G}\,{=}\,128^3$ grid results in numerous empty cells as shown by the top row. Each panel shows the count of sampling particles (just central particles at the voxel centres for this first case) within the cells. Instead of just transforming the voxel centre coordinates, we can instead implement a more abundant sampling process where we fill each voxel with $N_\text{p}$ randomly positioned sampling particles. This is shown in the second row where we use $N_\text{p}\,{=}\,5$ sampling particle per map voxel. This improves the empty cell problem but does not completely resolve it. A further simple improvement can be made by lowering the resolution of the Cartesian grid as done in the final row, halving the resolution to a $n_\text{G}\,{=}\,64^3$ grid. This results in a completely covered field and will avoid introducing additional ringing in a Fourier transform. 

The disadvantage in decreasing the Cartesian grid resolution, as in the final row of \autoref{fig:MonteCarloDemo}, is it artificially reduces the cutoff frequency beyond which small-scale (large-$k$) Fourier modes are inaccessible. The Nyquist frequency of the grid is defined as $k_\text{Nyq}\,{=}\,\pi/H$, where $H\,{=}\,l/n_\text{G}$ is the cell width of the grid, and $l$ is the grid length down one side in $h^{-1}\text{Mpc}$. If the underlying fluctuations contain power at $k\,{>}\,k_\text{Nyq}$, which we know to be the case with cosmological perturbations (even if damped by a low-aperture telescope), then the resulting gridding guarantees a loss of fidelity. Thus restricting this is essential for regridding. The lost power beyond the Nyquist frequency and the distortion it can cause through aliasing, are discussed in detail in the latter sections. To ensure a completely sampled Cartesian grid therefore, a balance is sought between limiting the number of Monte Carlo sampling particles per voxel, to keep computational efficiency under control, and avoiding a final Cartesian grid with too low a resolution and an artificially limited Nyquist frequency. 

The general survey geometry in Cartesian space is displayed by \autoref{fig:MonteCarloDemo}, which demonstrates how the constant sky area through frequency, which results in a larger spatial area being covered at larger distances (lower frequencies), provides the truncated pyramid footprint. This highlights the imprecision of the assumption that the survey footprint is approximately cubical in Cartesian space, even for a relatively small (${\sim} 20^2\,\text{deg}^2$, $0.37\,{<}\,z\,{<}\,0.58$) survey, which is already close to being surpassed by pathfinder LIM experiments. As sky area increases further, wide-angle effects which make the assumption that the line-of-sight is in the same parallel direction for every pixel, is no longer valid. This is covered in the literature \citep{Castorina:2017inr,Blake:2018tou} and becomes a non-negligible issue for very wide surveys, complicating anisotropic effects such as redshift space distortions, in addition to distorting the volume. In this work, we assume these issues are sub-dominant and avoid simulating anisotropies such as redshift space distortions for simplicity, since a robust simulation would require a projection of the velocity field along the line-of-sight in the frequency direction, which will not remain consistently parallel along any Cartesian axis.

\subsection{Simulated intensity maps}\label{sec:SimulatedIM}

To validate the Monte-Carlo regridding process introduced in the previous sub-section, we construct a suite of simulations covering two different scenarios. To provide statistically significant validation, we demand a large number of realisations of each simulation version. For this reason, we use fast lognormal mocks as the raw input field. The exact cosmology and emission line properties are not overly important, since in this work we are only examining what effect the discretisation and regridding processes have. This can be evaluated by studying the power spectra of the regridded fields relative to the input \textit{truth}, which will be the main focus for the rest of the paper. For completeness, however, a brief overview of the assumed input cosmology and lognormal generation process is provided in \appref{sec:FidCosmo}. We ensure the input starting simulation, $\delta_0$, has a high resolution ($n_0\,{=}\,512^3$ cubic voxels) relative to the subsequent sky maps and resampled Cartesian grid to mimic the reality of mapping a continuous field. 

The suite of simulations used for this work are generated by following the below presciption:
\begin{enumerate}[wide, labelwidth=!, labelindent=0pt]
    \item The size of the simulated survey is chosen in terms of its span in R.A, Dec and frequency (or redshift). From this, we compute the size of the Cartesian grid in which the survey footprint is entirely enclosed.\newline

    \item A lognormal mock $\delta_0$ is generated on the $n_0\,{=}\,512^3$ input grid. Every cell in this input grid has the correct Cartesian coordinates relative to an observer at the origin to be consistent with the survey footprint. We align the mock such that the line-of-sight through the centre of the survey is parallel with the Cartesian z-axis.\newline
    
    \item We construct a sky map, or lightcone\footnote{We use the term \textit{lightcone} loosely here, since there is no cosmological evolution within it. Instead, it is simply cutting the conical (or truncated pyramid) shape of the survey footprint from the input lognormal mock which has a consistent cosmology for a constant redshift $z_\text{eff}$, defined as the redshift for the median frequency in the survey.}, from the input mock, using \texttt{HEALPix} pixels at each frequency channel of the mock survey. The size of pixels and frequency channels are chosen to emulate realistic surveys or test certain scenarios. This involves an inverted Monte Carlo integration process, similar to that which we implement for the regridding to Cartesian space where all particles are transformed from Cartesian to sky coordinates and placed within the nearest (R.A., Dec, $\nu$) map voxel. We ensure each map voxel is well-sampled. Distributing sampling particles into sky coordinates to build a map is not dissimilar from real LIM map-making where calibrated time-ordered data stamps are collated into sky voxels. This can provide another source of aliasing which we discuss later.\newline
    
    \item From the sky maps, we execute the sampling process (outlined in \secref{sec:MoneteCarlosummary}) to provide the regridded field $\delta_\text{G}$ from which we can study the changes relative to the input field $\delta_0$. For the final regridding, a choice of resolution (number of FFT cells $n_\text{G}$) is made which we initially set to approximately double the size of the \texttt{HEALPix} sky voxels to ensure good sampling. Later in the paper, we revist the choice of $n_\text{G}$, increasing it so the Nyquist frequency of the analysis is pushed to higher-$k$.
\end{enumerate}

\begin{table}
    \setlength{\tabcolsep}{3.4pt}
	\centering
	\begin{tabular}{lccc} 
		\multicolumn{2}{c}{\textbf{Parameters}} & \multicolumn{2}{c}{\textbf{Simulation version}} \\

            \toprule
		& & Cubic voxels & Fine channel \\
		\toprule
        Sky area (approx.) [deg$^2$] & $A_\mathrm{sky}$ & 400 & 400 \\
        Bandwidth [MHz] & $\nu_\text{min}$ &  925 & 900 \\
         & $\nu_\text{max}$ & 1063 & 1100 \\
        \texttt{HEALPix} resolution & \texttt{nside} & 256 & 256 \\
        Channel width [MHz] & $\delta\nu$ & 1.17 & 0.5 \\
    	\midrule
        Redshift range & $z_\text{min}$ & 0.336 & 0.291 \\
         & $z_\text{max}$ & 0.535 & 0.578 \\
        Central redshift & $z_\text{eff}$ & 0.435 & 0.435 \\

        \texttt{HEALPix} pixel size [deg] & $\theta_\mathrm{pix}$ & 0.229 & 0.229 \\
        \# of filled $\texttt{HEALPix}$ pixels & $n_\text{pix}$ & 7182 & 7182 \\
        \# of freq. channels & $n_\nu$ & 118 & 400 \\
        Dish diameter [m] & $D_\mathrm{dish}$ & 15 & 15 \\
        Beam size (at $z_\text{eff}$) [deg] & $\theta_\mathrm{FWHM}$ & 1.15$^*$ & 1.15$^*$ \\
        
        \midrule

        Grid size [$h^{-1}\text{Mpc}$] & $l_\text{x},l_\text{y},l_\text{z}$ & 544, 547, 545 & 576, 579, 726\\
        \# of cells (input) & $n_{0\text{x}},n_{0\text{y}},n_{0\text{z}}$ & 512, 512, 512 & 512, 512, 512 \\
        \# of cells (FFT) & $n_{\text{G}\text{x}},n_{\text{G}\text{y}},n_{\text{G}\text{z}}$ & 59, 59, 59 & 59, 59, 200  \\
        \# of sampling particles & $N_{\text{p}}$ & 10 & 10  \\
        \bottomrule
	\end{tabular}
    \caption{Specifications for simulated data. Both simulation versions have 100 different realisations which we average results over. We chose to emulate a generic \hi\ 21cm LIM survey centred at $1\,\text{GHz}$, corresponding to a central \textit{effective} redshift of $z_\text{eff}\,{=}\,0.435$. A 3D grid size is calculated to fully enclose the survey footprint which defines the size of the input lognormal simulations. The final Cartesian grid has these same dimensions ($l_\text{x}$, $l_\text{y}$, $l_\text{z}$) but much broader cells ($n_{0\text{x}},n_{0\text{y}},n_{0\text{z}} \rightarrow n_{\text{G}\text{x}},n_{\text{G}\text{y}},n_{\text{G}\text{z}}$) initially chosen to approximately double the size of the \texttt{HEALPix} voxels. $^*$Note that in some cases, which we clearly identify, we do not include the beam, equivalently setting $\theta_\text{FWHM}\,{=}\,0$.}
    \label{tab:SurveyTable}
\end{table}

\begin{figure*}
    \centering
    \includegraphics[width=1\linewidth]{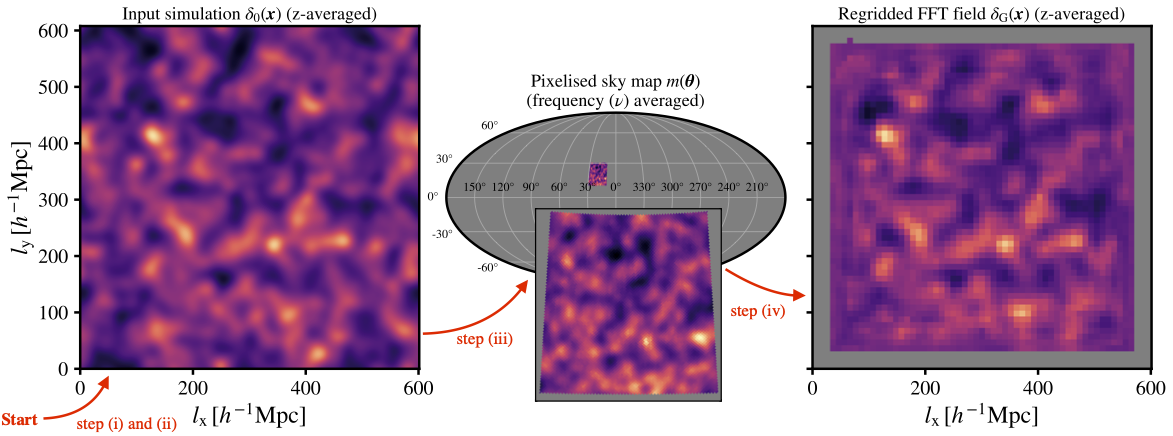}
    \caption{Mapping and regridding steps for one realisation of our simulated sky surveys covering $10^\circ\,{<}\,\text{R.A.}\,{<}30^\circ$ and $10^\circ\,{<}\,\text{Dec.}\,{<}30^\circ$. The red flowchart arrows and text indicate which steps in the simulation process (outlined near the beginning of \secref{sec:SimulatedIM}) are being performed. \textit{Left}-panel shows the starting simulation; a lognormal realisation generated from a power spectrum model. The size of the input field is determined so that the grid encloses the survey footprint, with some border padding to allow for non-local particle interpolation. \textit{Middle}-panel shows the full map in sky coordinates produced by sampling $\delta_0$ into \texttt{HEALPix} pixels and frequency channels with chosen resolutions $\texttt{nside}$ and $\delta\nu$. The middle-panel inset zooms in on the survey footprint showing the $20\,{\times}\,20\,\text{deg}^2$ field, also in sky coordinates. \textit{Right}-panel shows the final Cartesian grid resampled from the sky voxels with a chosen interpolation scheme for the sampling particles (we used nearest grid point assignment for this demonstration).}
    \label{fig:SimDemo}
\end{figure*}

\noindent \autoref{tab:SurveyTable} summarises the two simulation versions we adopt in this work. We begin with the \textit{Cubic voxels} simulation version which we artificially design to have approximately cubic dimensions for the voxels and Cartesian cells in the final grid $\delta_\text{G}$. This is to provide an \textit{isotropic} Nyquist frequency, to simplify the evaluation of aliasing effects. In reality, this may not be the case for real LIM experiments which can have very different resolutions along the line-of-sight (through frequency), relative to their broad angular pixels. We explore the consequences of this anisotropic Nyquist frequency in the other \textit{Fine channel} simulation version.

The choices made for the simulated surveys should not bear a huge influence on the generality of our results and conclusions. Nevertheless, we have chosen to approximately emulate a pathfinder 21cm intensity mapping experiment at radio wavelengths such as MeerKLASS \citep{MeerKLASS:2017vgf}. The $\texttt{nside}\,{=}\,256$ resolution of the map pixels we adopt throughout is approximately consistent with the $0.3\,\text{deg}$ MeerKLASS pixel size \citep{Wang:2020lkn}, approximately 1/3 of the beam size. \textcolor{black}{We also do not expect the choice of using a \texttt{HEALPix} pixelisation scheme to have an impact on the generality of our results. The current calibration pipeline for MeerKAT pilot data \citep{Wang:2020lkn} uses the Zenith Equal Area
(ZEA) map projection method \citep{Astropy:2013muo}. Thus we anticipate imminent validation of different pixelisation schemes which preceded the same Monte Carlo sampling techniques we outline in this work.}

The $20\,{\times}\,20\,\text{deg}$ survey size is arbitrarily chosen, but again is close to the current size of MeerKLASS pilot survey observations. This creates $7182$ filled sky pixels covering an approximately square patch. To generate cubic voxels we chose the frequency range and number of channels such that they span approximately the same depth $l_\text{z}$ as the spatial span in $l_\text{x}$ and $l_\text{y}$ on the Cartesian grid. We can also implement a smoothing to the field perpendicular to the line-of-sight to emulate the beam profile of the telescope. For this, we use a Gaussian profile assuming the dish size for the Square Kilometre Array Observatory \citep{Bacon:2018dui}, MeerKAT's successor, which has $D_\text{dish}\,{=}\,15\,\text{m}$. This defines the FWHM of the Gaussian beam profile in radians from $\theta_{\mathrm{FWHM}}\,{\approx}\, c / \nu_\text{eff} D_{\mathrm{dish}}$, where $\nu_\text{eff}$ is the median frequency of the survey, hence we assume a frequency-independent (constant size) beam.

\autoref{fig:SimDemo} shows a realisation from the \textit{Cubic voxels} simulation version. The left panel shows the input mock $\delta_0(\boldsymbol{x})$ averaged along the z-axis. This high-resolution input aims to emulate the continuous line emission field $\delta T_\text{LIM}(\boldsymbol{x})$ observed by the telescope. The middle panel shows the \texttt{HEALPix} pixelised sky map $m(\boldsymbol{\theta})$ averaged along $\nu$. The right panel shows the final regridded field $\delta_\text{G}(\boldsymbol{x})$ averaged along z.

When defining the size of the input Cartesian grid given the chosen survey size, we add some zero-padding around the borders. This is to allow for higher-order interpolation schemes (investigated later) in the map-to-grid resampling step, where particle assignment can be non-local, hence voxels around the edge will generate sampling particles outside the original footprint. This is seen in the final $\delta_\text{G}$ panel where there are some empty pixels due to the zero-order NGP assignment currently adopted. For higher-order schemes, this space becomes more filled. 

The Monte-Carlo resampling technique appears to be working by-eye. Original structure from the input $\delta_0$ can be seen in the final $\delta_\text{G}$ grid, albeit at a lower resolution. The edges of $\delta_\text{G}$ appear to show a greater loss in fidelity but this is merely a plotting effect caused by the averaging along the line-of-sight axis. Since the sky map $m$ cuts a conical (or truncated pyramid) footprint from the input $\delta_0$, there will be more spatial coverage at higher-z. Closer to the edges of the z-averaged map in $\delta_\text{G}$ there will be fewer slices covering the larger spatial separations, hence a lower sampling of the input $\delta_0$ and an apparent poor recovery of its structure.

\subsection{Biased power spectra from mapping and regridding}\label{sec:Pkbiasalias}

As briefly discussed in the introduction, previous pathfinder LIM survey analysis often avoided resampling the sky maps altogether by assuming the maps are approximately rectangular in Cartesian space. Before assessing the performance of our resampled maps, we briefly comment on the accuracy of this assumption for power spectrum measurements, to serve as motivation. We emulated the simplified approach by assuming that the $\nu$-stacked \texttt{HEALPix} maps from our $20\,{\times}\,20\,\text{deg}^2$ Cubic voxels simulation formed a rectangle in Cartesian space, with equal-sized voxels throughout. We assigned comoving dimensions to the edges of this rectangle, calculated with $l_\text{x}\,{=}\,d_\text{c}(z_\text{eff})\delta\theta_\text{RA}$, where $\delta\theta_\text{RA}\,{=}\,20\,\text{deg}$ (in radians). $l_\text{y}$ is calculated similarly but for Declination. The line-of-sight length is given by $l_\text{z}\,{=}\,d_\text{c}(z_\text{max})\,{-}\,d_\text{c}(z_\text{min})$. We then Fourier transform the sky map with these dimensions and measure the power spectrum. We found this approach caused a ${>}20\%$ bias across all scales relative to the input model and was most pronounced at large scales (small-$k$) reaching ${\sim}30\%$. The discrepancy increases at larger scales due to the true conical geometry of the survey, which means the approximation that separations on the sky are equivalent to separations in Cartesian space, is inaccurate. Whilst this approximation has been sufficient for many surveys with small sky coverage and low signal-to-noise, future surveys will require a more accurate approach, which is the motivation behind our resampling techniques.

The regridding of the LIM simulations from sky coordinates onto a Cartesian grid appears to be working well by eye, as seen by \autoref{fig:SimDemo}. However, the true test of performance will come from a statistical analysis of the field relative to a model with the same fiducial cosmology used to generate the input field $\delta_0$. The model we compare to therefore follows the input power spectrum discussed in \appref{sec:FidCosmo} and is interpolated over the same grid of modes as $\tilde{\delta}_\text{G}(\boldsymbol{k})$;
\begin{equation}\label{eq:Pkmod}
    P^{\prime}_\text{mod}(\boldsymbol{k},z) = \Big[b^2(z) \overline{T}^2(z)P_\text{m}(\boldsymbol{k},z) \exp \left(-k_\perp^2 R_{\text {beam}}^2(z)\right)\Big]*W_\text{G}(\boldsymbol{k})\,,
\end{equation}
where $P_\text{m}(\boldsymbol{k},z)$ is the matter power spectrum for the fiducial cosmology at redshift $z$. We include the option of modelling the telescope's beam which damps perpendicular modes $k_\perp$. For the frequency-independent Gaussian beam, this damping is modulated by the beam size $R_\text{beam}(z_\text{eff})\,{=}\,d_\text{c}(z_\text{eff})\theta_\text{FWHM}(z_\text{eff})/2\sqrt{2\ln2}$. We examine the effects of the beam and discuss further potential effects from more realistic, complex beam patterns later in the paper. The $^\prime$ notation on $P^{\prime}_\text{mod}$ in \autoref{eq:Pkmod} is to notify the different iterations of model improvements which will be useful later when extending this base model.

The $*\,W_\text{G}(\boldsymbol{k})$ part of \autoref{eq:Pkmod} represents the convolution of the model with the survey window function, and optional optimal weighting. This is necessary since mapping the input field onto the survey footprint leaves empty pixels in the final regridded field $\delta_\text{G}$ outside the survey footprint, demonstrated by the white space in the boarders of \autoref{fig:MonteCarloDemo}. To correct for this in the model, we must convolve with a survey window function which we simply define as $W_\text{G}(\boldsymbol{x})\,{=}\,1$ for every non-zero cell on the grid covered by the survey footprint, and zero otherwise. More complex survey window functions can be implemented to account for non-uniform survey completion. Some apodization (or smoothing) of the footprint edges may help any discontinuous edge effects in the mask but is not something we investigate in detail in this work. Convolving the model with this binary window function, and normalising by it and any optional survey weighting can be expressed for a generic model power spectrum $P_\text{mod}$ with
\begin{equation}\label{eq:Pkmod_windowconv}
    P_\text{mod}(\boldsymbol{k}) * W_\text{G}(\boldsymbol{k}) = \frac{\sum_i P_\text{mod}(\boldsymbol{k}^\prime_i) |\tilde{W}_\text{G}(\boldsymbol{k}-\boldsymbol{k}_i^\prime)|^2}{\sum w^2(\boldsymbol{x})W^2_\text{G}(\boldsymbol{x})}\,.
\end{equation}
The model from \autoref{eq:Pkmod} is averaged into the same spherical shell bins as the power estimation for the simulations, which we chose to be linear in $k$.

\begin{figure}
    \centering
    \includegraphics[width=1\linewidth]{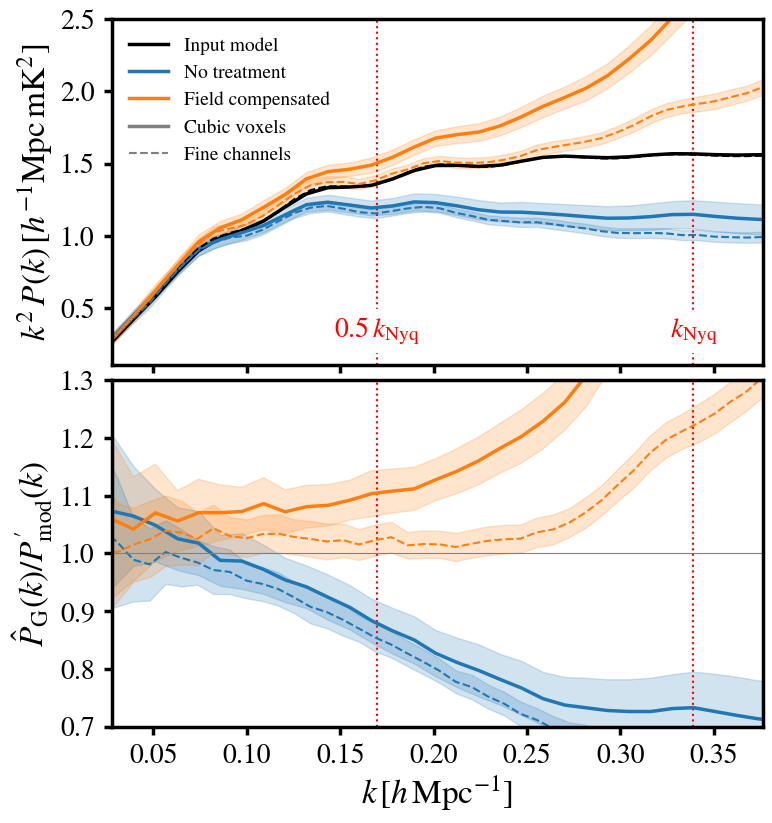}
    \caption{Accuracy of the measured power spectrum of the regridded field relative to the model (black lines) for the two simulation versions (Cubic voxels and Fine channels). We show the case where no compensation for the NGP interpolation is applied to the regridded field (blue lines) and the case where compensation is applied using \autoref{eq:Compensated} and \autoref{eq:W_ngp} (orange lines). We do not include the effects of the telescope beam in these results.  Measured power spectra $\hat{P}_\text{G}$ are averaged over 100 simulation realisations. The $1\sigma$ scatter from the 100 simulations is shown by the shaded regions.}
    \label{fig:Pk_ngp_bias}
\end{figure}

\autoref{fig:Pk_ngp_bias} shows the power spectrum estimation of the regridded field for the Cubic voxels simulation (solid line), initially with no beam i.e $R_\text{beam}\,{=}\,0$. For the simulation results, we average over the 100 realisations for each version. We found this was enough realisations to reach converged results for all tests in this paper. The shaded regions show the $1\sigma$ range over the realisations to indicate the scatter from the simulations. Since the main focus in this work is optimising \textit{accuracy}, we do not show these uncertainty regions in subsequent plots, since their size is driven by the survey size choice and not any analysis treatments we implement. The red dotted vertical lines (also in subsequent similar plots in the paper) represent the Nyquist frequency $k_\text{Nyq}$ and $0.5k_\text{Nyq}$ of the Cartesian grid for the perpendicular x and y directions. For the Cubic voxels simulation, the z-direction Nyquist frequency is also equal to the x and y directions by design. For the Fine channels version, the z-direction has a higher Nyquist frequency $k_\text{Nyq}\,{=}\,0.865\,h\,\text{Mpc}^{-1}$, due to the increased resolution along the line-of-sight.

The model (shown by the black lines in the top-panel of \autoref{fig:Pk_ngp_bias}) is directly compared to in the bottom panel, where perfect agreement would see $P(k)/P_\text{mod}(k)\,{=}\,1$. There is poor agreement for the blue solid line which has no further treatment than what we have already discussed. We discuss the results from the \textit{Fine channels} version (dashed lines) shortly. The modelling should be improved by recognising that in the construction of the $\delta_\text{G}(\boldsymbol{x})$ field, we performed an interpolation of the sampling particles using a NGP assignment scheme. As identified in \citet{Jing:2004fq}, assigning a particle distribution that traces a continuous tracer field $\rho(\boldsymbol{x})$ onto a grid $\rho(\boldsymbol{x}_\text{G})$ is equivalent to a convolution of the continuous field with a mass assignment scheme window function $W_\text{mas}(\boldsymbol{x})$. In Fourier-space this convolution becomes a multiplication, thus our LIM data resampled onto the grid $\delta_\text{G}$ can have the leading-order effects corrected by simply dividing the field by the Fourier transform of the window function $\tilde{W}_\text{mas}(\boldsymbol{k})$ \citep{Sefusatti:2015aex}, i.e.
\begin{equation}\label{eq:Compensated}
    \tilde{\delta}_\text{G}(\boldsymbol{k}) \rightarrow \frac{\tilde{\delta}_\text{G}(\boldsymbol{k})}{\tilde{W}_\text{mas}(\boldsymbol{k})}\,.
\end{equation}
For the NGP assignment scheme that we currently use, we are simply discretising particles into cells of size $H\,{=}\,l/n_\text{G}$. This is modelled by a top-hat function which the Fourier transform of is a sinc function, $\text{sinc}(x)\,{=}\,\sin(x)/x$. For a 3D field, these window functions remain separable i.e. $W(\boldsymbol{x})\,{=}\,W(x)W(y)W(z)$ \citep{HockneyEastwood} and the interpolation window function for NGP is given by
\begin{equation}\label{eq:W_ngp}
    \tilde{W}_{\rm ngp}(\boldsymbol{k})= \text{sinc}\left(\frac{k_\text{x} H_\text{x}}{2}\right)
    \text{sinc}\left(\frac{k_\text{y} H_\text{y}}{2}\right)
    \text{sinc}\left(\frac{k_\text{z} H_\text{z}}{2}\right)\,.
\end{equation}
The orange results in \autoref{fig:Pk_ngp_bias} show the results from applying this compensation directly to the field defined by \autoref{eq:Compensated}. We see how this is providing some correction for the damping due to the NGP interpolation (orange solid line) and now power is enhancing as the Nyquist frequency is approached. This shows the impact of aliased power caused by small-scale separations between sampling particles not resolved by the cells in $\delta_\text{G}$. Aliasing will still be present without the compensation in \autoref{eq:Compensated} (blue lines), but the results are overly dominated by the smoothing from the particle assignment. The following section is devoted to how further improvements can be made both to mitigate this aliasing and to also include modelling for prior discretisation to the field in the map-making stage i.e. performing $\delta_0(\boldsymbol{x})\rightarrow m(\boldsymbol{\theta},\nu)$.

Before exploring further modelling and improved interpolation techniques, we consider the case where the LIM frequency channels have better resolution than the pixels, which will emulate a realistic situation for many LIM sureys. We use the \textit{Fine channels} simulation version (see \autoref{tab:SurveyTable}), where the number of frequency channels is more than trebled, increasing the maximum $k_\parallel$ accessible. We also increase the redshift range probed which is equivalent to increasing the depth of the survey in comoving space. We do all this in such a way to keep the central redshift $z_\text{eff}$ unchanged to aide result comparison. This departs from the simple case of having approximately cubic voxels and cells. Now there will be a different Nyquist frequency depending on the axis direction which will have interesting effects for aliasing. 

\begin{figure}
    \centering
    \includegraphics[width=1\linewidth]{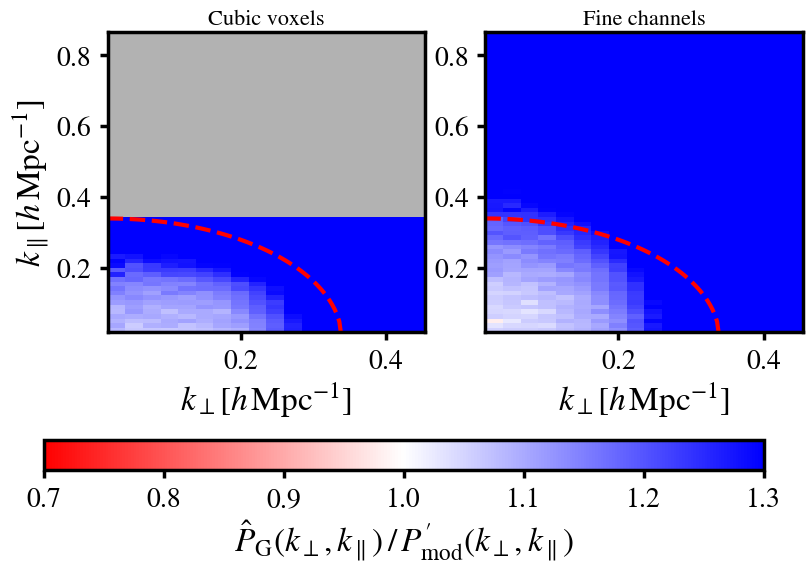}
    \caption{Cylindrical ($k_\perp, k_\parallel$) power spectrum comparison to model for the \textit{Cubic voxels} (left panel) and \textit{Fine channel} (right panel) simulations. Blue regions represent overly enhanced power relative to the true model driven by aliasing. The red dashed line shows the same \textit{minimum} Nyquist frequency as in \autoref{fig:Pk_ngp_bias} calculated from the cell resolution along one of the dimensions in the perpendicular directions. Colour-bar range is fixed to avoid saturation.}
    \label{fig:Pk2D_alias}
\end{figure}

The results for the Fine channel simulation are shown by the dashed lines in \autoref{fig:Pk_ngp_bias}. For the case where we compensate for the NGP interpolation (orange dashed line), there is less enhanced power compared to the Cubic voxels (orange solid line). This is because, there will be less aliasing along the line of sight when there is higher frequency resolution, hence aliased contributions to the spherically averaged modes will be suppressed. Analysing the 2D \textit{cylindrically}-averaged power spectrum decomposed into $(k_\perp, k_\parallel)$ modes demonstrates this point. \autoref{fig:Pk2D_alias} shows the 2D power for the cubic voxels (left panel) and fine channel case (right panel). We draw the same Nyquist frequency (red dashed line) from \autoref{fig:Pk_ngp_bias} for reference. This is defined as the minimum Nyquist frequency which comes from the perpendicular dimensions, hence is the same for both panels. The grey area for the Cubic voxels is an area of $k$-space inaccesible by the grid resolution, but can be accessed by the Fine channels simulation. The blue regions show the area of $k$-space where aliasing dominates. For the cubic voxels, power enhances relatively isotropically as $|\boldsymbol{k}|\,{=}\,\sqrt{k_\parallel^2\,{+}\,k_\perp^2}$ approaches the Nyquist frequency. For the fine channels however, aliasing does not dominate along the parallel axis until much higher $k_\parallel$. Therefore, aliasing in the spherically averaged modes around the reference scale (the minimum Nyquist frequency $k_\text{Nyq}$) will wash out as seen in the 1D power for the Cubic voxels in \autoref{fig:Pk_ngp_bias}. Aliasing remains present in regions of $k$-space even with fine channel resolution and thus we still require an extension on the techniques we have used so far, which we explore in the following section. Before that however, we consider the impact from the telescope beam.

\subsubsection{Why aliasing remains even with a large instrumental beam}

To complete this section, we consider the impact from the telescope's instrumental beam. So far the results presented have been for the case with no perpendicular smoothing to the field caused by the beam. One might naively assume that the smoothing will completely suppress aliasing effects, however, this is not necessarily the case since smoothing from the beam comes before the sampling particle assignments. Therefore, any aliasing effects originating from particle assignment (or time stamp binning in map-making), will still inject aliased power, artificially enhancing the beam-damped power spectrum. Since fluctuations still exist beyond the Nyquist frequency, albeit damped by the beam, the power spectrum will still be distorted by aliasing. \textcolor{black}{For this reason, we also expect aliasing to remain in the presence of \textit{any} smoothing caused by instrumentation. Higher-frequency LIM experiments e.g. SPHEREx \citep{SPHEREx:2014bgr}, will have a less precise spectral resolution, which can be characterised by a radial damping, similar to the beam but in an orthogonal direction to it. Hence, the effects of aliasing from resampling the field are likely to be an issue for any LIM experiment aiming for precision cosmology.}

\begin{figure}
    \centering
    \includegraphics[width=1\linewidth]{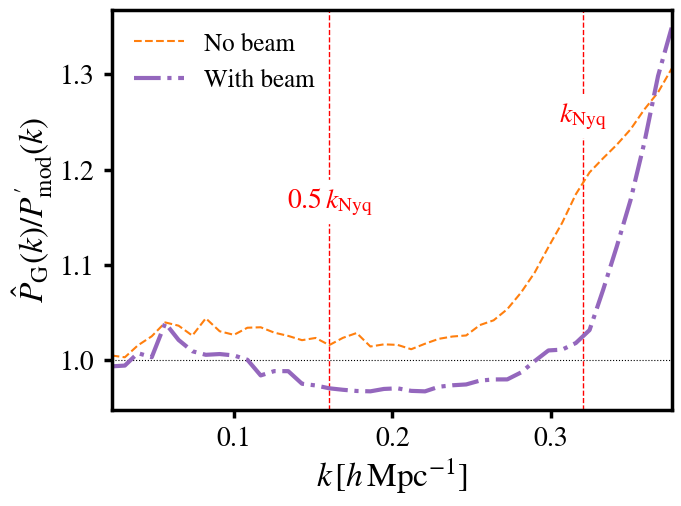}
    \caption{Accuracy of the measured power spectrum relative to the model for the case with and without smoothing effects from the telescope beam. Measured power spectra $\hat{P}_\text{G}$ are averaged over 100 simulation realisations.}
    \label{fig:Pk_nobeam}
\end{figure}

\autoref{fig:Pk_nobeam} shows the impact from the beam in the regridded power spectrum measurement. We smooth the input field $\delta_0$ with a $\theta_\text{FWHM}\,{=}\,1.15\,\text{deg}$ beam, equivalently $R_\text{beam}\,{=}\,9.6\,h^{-1}\text{Mpc}$. This is not a perfect emulation of reality as the beam will actually act perpendicularly in the curved sky which will not be perfectly perpendicular in the Cartesian frame. We discuss the interesting impact from the beam further in \appref{sec:BeamEffects}. Despite using an oversimplified Gaussian beam applied in the Cartesian frame, it provides a sufficient test to show that fields smoothed by a beam do not escape aliasing. Since we can trivially model the impact of this simple beam by including $R_\text{beam}$ in \autoref{eq:Pkmod}, we continue to implement it in the rest of the paper for some extra completeness, but defer a more robust implementation of it to future work, as also discussed in \appref{sec:BeamEffects}.

\section{Mitigating aliasing and mapping effects}\label{sec:MitigatingAliasingContributions}

In \secref{sec:RegridMonte} we introduced the Monte-Carlo resampling which achieved a decent accuracy on large scales but introduced some effects at smaller scales. Even after correcting for the smoothing to the field (\autoref{eq:Compensated}) from the interpolation of sampling particles, a discrepancy between power spectrum measurement and model remained. This will be caused by aliased power introduced by the sampling particles as well as the additional discretisation step from the prior \textit{voxelisation} into the sky maps. In this section we extend the modelling used in the previous section to approximate the effects from the sky mapping, with the aim of improving the accuracy of the final power spectrum estimation. Additionally, we explore some higher-order interpolation schemes for the Monte Carlo sampling particles beyond the simple nearest grid point (NGP) assignment scheme used by default so far, which we showed introduces aliased power.

\subsection{Modelling the sky mapping}

An exact model for the additional step of discretisation from mapping the ``continuous'' field $\delta_0$ into (R.A., Dec. $\nu$) voxels on $m(\boldsymbol{\theta},\nu)$, would require consideration of the curved sky and the spherical shell voxels that the temperature fluctuations are discretised into. However, following \citet{Blake:2019ddd}, by making a narrow-angle assumption, we can make the approximation that the \texttt{HEALPix} pixelisation and frequency channel binning are separable effects on $k_\perp$ and $k_\parallel$ respectively. 

For the angular pixelisation, we utilise the \texttt{HEALPix} pixel window function\footnote{\href{https://healpy.readthedocs.io/en/latest/generated/healpy.sphtfunc.pixwin.html}{healpy.readthedocs.io/en/latest/generated/healpy.sphtfunc.pixwin.html}} $W_\text{pix}(\ell)$ which describes the damping to the harmonic angular power spectrum $C_\ell \rightarrow C_\ell W^2_\text{pix}(\ell)$, given by
\begin{equation}\label{eq:Wpix}
    W_{\mathrm{pix}}^2(\ell)=\frac{4 \pi}{2 \ell+1} \sum_{m=-\ell}^{\ell}\left|w_{\ell m}\right|^2\,,
\end{equation}
where
\begin{equation}
    w_{\ell m}=\int_{\text{pixel}} \text{d}\Omega \, Y_{\ell m}(\Omega)\,,
\end{equation}
is the spherical harmonic transform of a pixel in terms of the spherical harmonic functions $Y_{\ell m}$. For the final Cartesian grid $\delta_\text{G}(\boldsymbol{x})$, we assume $\ell\,{\propto}\,k_\perp$ and model the damping caused by the pixelisation as $P(\boldsymbol{k})\rightarrow P(\boldsymbol{k})W_\text{pix}^2(\ell_\text{eff})$, where $\ell_\text{eff}\,{=}\,k_\perp/d_\text{c}(z_\text{eff})$, and $d_\text{c}$ is the comoving distance to the central redshift of the survey. Thus the damping due to the angular pixelisation can be expressed with
\begin{equation}\label{eq:Bpix}
    \tilde{B}_\text{pix}(\boldsymbol{k}) = W_{\mathrm{pix}}(k_\perp/d_\text{c}(z_\text{eff}))
\end{equation}
There is also a smoothing effect along the line-of-sight from the sky mapping due to the discrete frequency channels of the LIM survey. This acts as a top hat function along the frequency axis, which we approximate as being consistently aligned along $k_\parallel$. For channels with depth $\Delta r_\text{chan}$, this is modelled with a sinc function in Fourier space given by
\begin{equation}\label{eq:Bchan}
    \tilde{B}_{\text {chan}}(\boldsymbol{k})=\text{sinc} \left(\frac{k_\parallel \Delta r_\text{chan}(\boldsymbol{x})}{2}\right)\,.
\end{equation}

\subsubsection{Additional aliasing from sky mapping}

During the creation of the \texttt{HEALPix} sky maps $m(\boldsymbol{\theta},\nu)$, aliased frequencies will be inserted into the sky map, from the default NGP assignment in the map making. Since the map undergoes further regridding onto the Cartesian field, the original field in which these map-based effects arise is not accessible, thus we opt to apply a further correction in our modelling instead. \textcolor{black}{There is scope to address this problem at the map-making stage with higher-order assignment of the time-ordered data into \texttt{HEALPix} (or an alternative scheme's) map pixels. Due to the extra requirement of simulating time-ordered data to test this, we leave this investigation to future work.} To mitigate the bias we \textcolor{black}{instead} follow an approach closely aligned to that outlined in \citet{Jing:2004fq} (see also \citet{Blake:2010xz} for an application with real data), whereby the power spectrum is summed over modes displaced by integers of $2k_\text{Nyq}$ to correct for this discretisation at sky map level. This requires the grid onto which the model is interpolated be representative of the sky voxel sizes. Once again, this is not perfectly possible due to sky curvature but the pixel widths and channel depths can be approximated which is simply done by choosing the number of Fourier cells $(n_\text{Gx}, n_\text{Gy}, n_\text{Gz})$ to equal the number of map pixels and frequency channels. If the overall Fourier grid size is calculated in a way that is consistent with the Cartesian size of the map space, the FFT cells will be approximate to the voxel sizes. We can then perform a summation of displaced modes onto the model which should approximately suppress the map voxelisation effects. This has the added benefit of sampling onto a more resolved grid than before, where the cell size was approximately double the map voxel sizes. Providing we avoid the empty cell problem as presented by \autoref{fig:MonteCarloDemo}, this will have the beneficial effect of not artificially decreasing the Nyquist frequency below that which is defined by the map voxel sizes. This keeps discrepant effects contained to smaller scales which can be disregarded in a final analysis.

\begin{figure}
    \centering
    \includegraphics[width=1\linewidth]{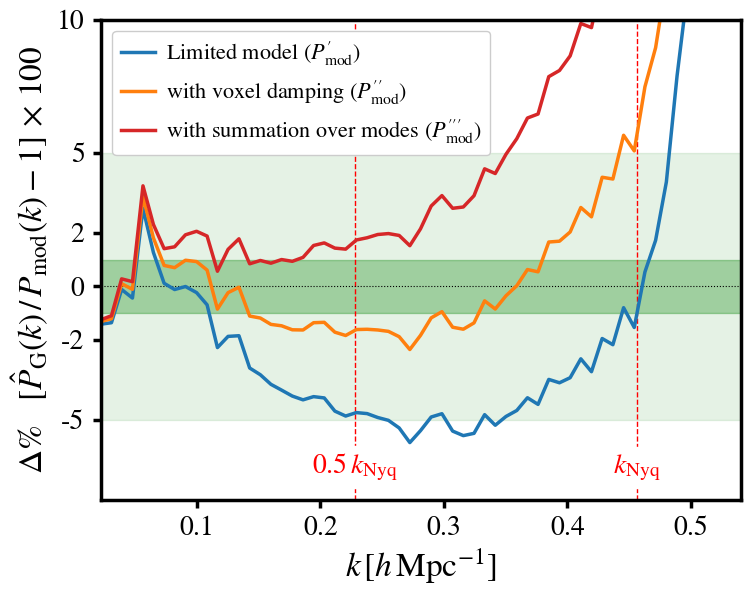}
    \caption{Percent accuracy of the measured power spectrum relative to increasingly sophisticated models ($P^\prime_\text{mod}$ \autoref{eq:Pkmod}), ($P^{\prime\prime}_\text{mod}$ \autoref{eq:Pkmod_vox}) and ($P^{\prime\prime\prime}_\text{mod}$ \autoref{eq:sumovermodes}). Green shaded areas show the sub-5\% and 1\% accuracy regions. Measured power spectra $\hat{P}_\text{G}$ are averaged over 100 simulation realisations. Blue line shows the limited model from \secref{sec:RegridMonte}. The orange line includes damping from the discretisation of the field into the map pixels and frequency channels. The red line further includes the summation over modes.}
    \label{fig:Pk_modelpix}
\end{figure}

Any aliasing caused by the Monte Carlo sampling particles onto the final Cartesian grid can still be corrected for at the field level, so this approach partitions the effects with the sky mapping discretisation effects handled in the modelling, and Cartesian regridding effects addressed at field-level. We summarise the final modelling approach with the below two steps;
\begin{enumerate}[wide, labelwidth=!, labelindent=0pt]
    \item The first extension to the original $P^\prime_\text{mod}$ base model (\autoref{eq:Pkmod}) is to include damping from the map voxel discretisation, achieved with the smoothing functions in \autoref{eq:Bpix} and \autoref{eq:Bchan};
    \begin{equation}\label{eq:Pkmod_vox}
    \begin{aligned}
        P_\text{mod}^{\prime\prime}(\boldsymbol{k},z) = \Big[b&^2(z) \overline{T}^2(z)P_\text{m}(\boldsymbol{k},z) \exp \left(-k_\perp^2 R_{\text {beam}}^2(z)\right)\\
        &\times 
        \tilde{B}_\text{pix}^2(\boldsymbol{k})\,\tilde{B}^2_{\text {chan}}(\boldsymbol{k})\Big]\,*W_\text{G}(\boldsymbol{k}).
    \end{aligned}
    \end{equation}
    \\
    As before, this model is convolved with the survey window functions following \autoref{eq:Pkmod_windowconv}.
    The results from this relative to the original model are shown in \autoref{fig:Pk_modelpix} (orange line).
    \\
    \item The final extension applied to the model, after convolution with the survey window function, is the correction to mitigate any bias from the NGP assignment into sky map voxels. For this we adopt a correction following \citep{Jing:2004fq} which can equivalently be applied with the summation over modes;
    \begin{equation}\label{eq:sumovermodes}
        P_\text{mod}^{\prime\prime\prime}(\boldsymbol{k}) = \sum_{\boldsymbol{n}} \tilde{W}^2_\text{ngp}(\boldsymbol{k}\,{+}\,2\boldsymbol{n}k_\text{Nyq})\,P^{\prime\prime}_\text{mod}(\boldsymbol{k}\,{+}\,2\boldsymbol{n}k_\text{Nyq})\,, 
    \end{equation}
    where $\boldsymbol{n}$ is a vector of integers, a summation over which displaces $\boldsymbol{k}$ by $2\boldsymbol{n}k_\text{Nyq}$ in all dimensions. In principle, this can extend indefinitely but it is usually sufficient to just consider the leading order displacement. We explored this and found no improvement when going beyond a $3^3$ grid of $\boldsymbol{n}\,{=}\,\{-1,0,1\}$. In \autoref{eq:sumovermodes}, the model requires interpolation onto a grid whose cells are a close approximation to the sky voxel size, so that the displaced Nyquist frequencies and NGP window function are emulating the sky map effects. 
    
\end{enumerate}

\noindent As before, the plotted model power spectra are all averaged into the same spherical shell bins as the power estimation for the simulations.

Results for the full model, including the summation over modes, are shown by the red line in \autoref{fig:Pk_modelpix}. There is no clear improvement by introducing the summation over modes, but this is still for the case where the zeroth order NGP assignment scheme is being used for the Monte Carlo sampling particles. Extending beyond this should mitigate the enhanced power shown by the red line which will be a result of residual aliasing, mostly from the Monte Carlo resampling.

\subsection{Higher-order mass assignment schemes}\label{sec:HighOrderMAS}

We have already shown in \secref{sec:Pkbiasalias} how the smoothing effects from the sampling particle assignments can be modelled by a convolution with the window function (\autoref{eq:W_ngp}) in Fourier space, which is simply a product of separable top-hat functions along each dimension. As shown in \citet{HockneyEastwood}, iteratively convolving with further top-hat functions is an effective approach for delivering higher-order interpolations to suppress aliasing. These higher-order routines generate 3D \textit{cloud} shapes and the portion of density from each shape that falls within a particular cell determines the value assigned to that cell from the initial mass introduced. An easier way to illustrate the interpolations is with 1D assignment function shapes. The first-order function, cloud-in-cell (CIC) is piecewise linear, as apposed to NGP which is piecewise constant. Beyond this, the triangular-shaped cloud (TSC) scheme becomes a continuous first derivative, with increasingly higher-order derivatives beyond this. Following \citet{Sefusatti:2015aex}, we test up to the third-order interpolation scheme now commonly referred to as the piecewise cubic spline (PCS).

\begin{figure}
    \centering
    \includegraphics[width=1\linewidth]{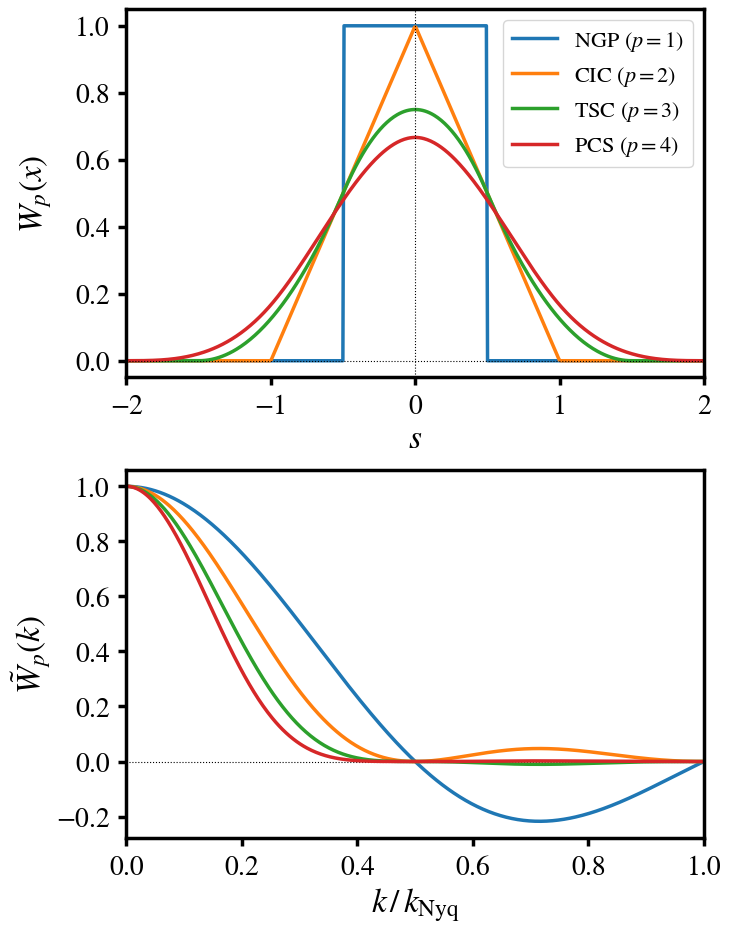}
    \caption{Window functions in 1D for the different mass assignment schemes considered in this work. Top panel shows the different interpolations in configuration space used to assign the particle intensities into cells as a function of the normalised distance from the cell $s$. The bottom panel shows the Fourier transform of the window functions.}
    \label{fig:interpolation_functions}
\end{figure}

We express the 1D assignment function shapes for the four schemes we consider below as a function of $s$, the normalised particle distance from a cell-centre $x^\text{G}_i$, i.e. $s\,{=}\,(x^\text{G}_i\,{-}\,x)/H$, where once again $H\,{=}\,l/n_\text{G}$ is defined as the cell size;
\begin{itemize}[leftmargin=*]
    \item Nearest Grid Point (NGP):
    \begin{equation*}
        W_{p{=}1}(s)= \begin{cases}1 & \text { for }|s|<\frac{1}{2} \\ 0 & \text { otherwise }\end{cases}    
    \end{equation*}
    \item Cloud-in-Cell (CIC):
    \begin{equation*}
        W_{p{=}2}(s)= \begin{cases}1-|s| & \text { for }|s|<1 \\ 0 & \text { otherwise }\end{cases}
    \end{equation*}
    \item Triangular-Shaped Cloud (TSC)
    \begin{equation*}
        W_{p{=}3}(s)= \begin{cases}\frac{3}{4}-s^2 & \text { for }|s|<\frac{1}{2} \\ \frac{1}{2}\left(\frac{3}{2}-|s|\right)^2 & \text { for } \frac{1}{2} \leqslant|s|<\frac{3}{2} \\ 0 & \text { otherwise }\end{cases}
    \end{equation*}
    \item Piecewise Cubic Spline (PCS)
    \begin{equation*}
        W_{p{=}4}(s)=\left\{\begin{array}{lc}
\frac{1}{6}\left(4-6 s^2+3|s|^3\right) & \text { for } 0 \leqslant|s|<1 \\
\frac{1}{6}(2-|s|)^3 & \text { for } 1 \leqslant|s|<2 \\
0 & \text { otherwise }
\end{array}\right.
    \end{equation*}
    
\end{itemize}
This describes the practical interpolation of the sampling particles onto the regridded field $\delta_\text{G}$. \autoref{fig:interpolation_functions} (top panel) displays these assignment functions in configuration space as a function of $s$, showing the increasing non-locality of the higher order functions, hence computational demand increases with $p$, since more neighboring cell assignments require calculation for each particle. In this work, we utilise the public \texttt{python} package \texttt{pmesh}\footnote{\href{https://rainwoodman.github.io/pmesh/index.html}{rainwoodman.github.io/pmesh/index.html}}, also used for other cosmological analysis codes such as \texttt{nbodykit}\footnote{\href{https://nbodykit.readthedocs.io/en/latest/index.html}{https://nbodykit.readthedocs.io/en/latest/index.html}} \citep{Hand:2017pqn}.
The window functions in Fourier space for each of these schemes is then simply the sinc functions raised to the power $p$ which depends on the choice of mass assignment scheme;
\begin{equation}\label{eq:W_mas}
    \tilde{W}_{p}(\boldsymbol{k})= \left[\text{sinc}\left(\frac{k_\text{x} H_\text{x}}{2}\right)
    \text{sinc}\left(\frac{k_\text{y} H_\text{y}}{2}\right)
    \text{sinc}\left(\frac{k_\text{z} H_\text{z}}{2}\right)\right]^p\,,
\end{equation}
where $p=\{1,2,3,4\}$ for NGP, CIC, TSC and PCS respectively. We display these window functions in the bottom panel of \autoref{fig:interpolation_functions} which shows the desired effect of making the assignment more local in Fourier-space and should suppress aliasing contributions. The price to pay for making the assignment more local in Fourier space is the configuration-space assignment is increasingly \textit{non}-local (as shown by the top panel), thus increasing the computational cost. Even more non-local interpolations can be considered e.g. using Daubechies wavelets \citep{Cui:2008fi}. Results in \citet{Hand:2017pqn} showed that despite their improved performance at smaller scales, some larger scale discrepancy is introduced, hence we do not investigate these schemes in this work.

In \autoref{fig:Pk_mas_comparison}, we show the accuracy of the higher-order interpolation scheme results (dashed lines), relative to the full model case $P^{\prime\prime\prime}_\text{mod}$ \autoref{eq:sumovermodes}. We discuss these results in more detail soon, but first, we introduce the additional field-level treatment of interlacing.

\subsubsection{Interlacing}

\begin{figure}
    \centering
    \includegraphics[width=1\linewidth]{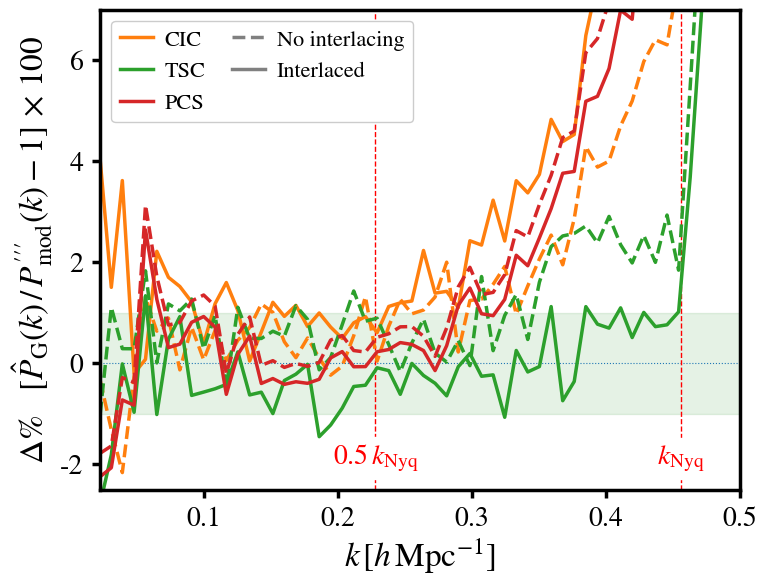}
    \caption{Percent accuracy of the measured power spectrum relative to full model $P_\text{mod}^{\prime\prime\prime}$ (\autoref{eq:sumovermodes}) for the higher-order mass assignment schemes, both with and without interlacing. The green shaded area shows the sub-1\% accuracy region. Measured power spectra $\hat{P}_\text{G}$ are averaged over 100 simulation realisations.}
    \label{fig:Pk_mas_comparison}
\end{figure}

\citet{Sefusatti:2015aex} investigated using interlaced FFT grids as a means for suppressing aliasing, as initially introduced in \citet{HockneyEastwood}. Interlacing involves interpolating onto a second FFT grid with the cell positions all shifted by half a cell size ($H/2$) in all directions, so the discrete Fourier transform of this second field will be given by
\begin{equation}
    \tilde{\delta}_2^\text{G}(\boldsymbol{k})=\sum_{\boldsymbol{x}}{\delta}_\text{G}(\boldsymbol{x} + H/2) \, w(\boldsymbol{x} + H/2) \exp (i \boldsymbol{k} \cdot \boldsymbol{x})\,.
\end{equation}
This is combined with the original unshifted field, from \autoref{eq:deltaG_k}, repeated here;
\begin{equation*}
    \tilde{\delta}_1^\text{G}(\boldsymbol{k})=\sum_{\boldsymbol{x}} \delta_\text{G}(\boldsymbol{x}) w(\boldsymbol{x}) \exp (i \boldsymbol{k} \cdot \boldsymbol{x})\,,
\end{equation*}
to provide the interlaced field which is the linear combination of the two, given by
\begin{equation}
    \tilde{\delta}_{\mathrm{G}}(\boldsymbol{k})=\frac{1}{2}\left[\tilde{\delta}_1^{\mathrm{G}}(\boldsymbol{k})+\tilde{\delta}_2^{\mathrm{G}}(\boldsymbol{k})\right]\,. 
\end{equation}
The interlacing technique is inspired by the ``butterfly'' operation which the FFT is based upon. As analytically shown in \citet{Sefusatti:2015aex}, the interlaced field will cancel out the contributions from the odd integer phases of the aliased modes which should include the leading order contributions.\newline 
\\
\noindent \autoref{fig:Pk_mas_comparison} presents the higher-order interpolation scheme results, both with and without interlaced fields. Relative to the full model case $P^{\prime\prime\prime}_\text{mod}$ (\autoref{eq:sumovermodes}). This shows how the enhanced interpolations are generally suppressing the aliasing and allowing a sub-percent agreement with the model power. 

There is a less hierarchical order of results in \autoref{fig:Pk_mas_comparison} than expected, and as seen in equivalent investigations for galaxy catalogues \citep[e.g.][]{Jing:2004fq,Sefusatti:2015aex,Hand:2017pqn}. In these studies, where a straightforward gridding of point-like galaxies is performed, the higher-order schemes deliver a clear improvement in accuracy. For our results however, we appear to have diminishing returns and it even appears that the TSC scheme performs better than the higher-order PCS at high-$k$ (although this is not the case at larger linear scales as we will show). We suspect that the lack of conclusive successive improvement with increasingly more sophisticated schemes is due to the complication we face from the additional discretisation step caused by the mapping into sky voxels. This introduces a NGP interpolation of the continuous field, which we then apply a further interpolation to in the Monte-Carlo resampling to the Cartesian grid. Whilst our approximate models for these effects are performing well, they will not untangle the complicated mix of particle assignments that we believe make the final higher-order interpolations less efficient at suppressing residual aliasing. We were able to test this hypothesis in a Cartesian-only investigation where the sky mapping step was done onto another Cartesian field, thus avoiding curved-sky complications. This toy case reproduced similar results where some small residual aliasing remains from the extra mapping step and there were still no clear hierarchical improvements as higher-order interpolation schemes were implemented. We leave further investigation to future work into whether an enhanced map-making process can be introduced that avoids this consequence and suppresses aliased power further. 

\begin{table}
    \centering
    \begin{tabular}{|c|c|c|}
         \textbf{Scheme} & \textbf{Interlacing} & \textbf{Average bias [\%]} \\
         \midrule
         NGP & False & 1.381 \\ 
         NGP & True & 0.431\\ 
         CIC & False & 0.528 \\ 
         CIC & True & 0.948\\ 
         TSC & False & 0.715\\ 
         TSC & True & 0.568\\ 
         PCS & False & 0.500\\ 
         PCS & True & 0.382 \\
        \toprule
    \end{tabular}
    \caption{Average percentage bias relative to model in results from \autoref{fig:Pk_mas_comparison} for scales $0.07\,{<}\,k\,{<}\,0.25\,h\,\text{Mpc}^{-1}$. For completeness, we include the NGP results not plotted in \autoref{fig:Pk_mas_comparison}. Percentages are the average of the \textit{absolute} bias i.e. $100\times|\hat{P}_\text{G}/P_\text{mod} - 1|$, to avoid positive and negative biases cancelling.}
    \label{tab:AverageBias}
\end{table}

The results from \autoref{fig:Pk_mas_comparison} sufficiently demonstrate how sub-percent accuracy for regridded LIM data is achievable far up to the Nyquist frequency of the Cartesian grid. Despite a better performance of TCS for high-$k$, on medium to large scales ($0.07\,{<}\,k\,{<}\,0.25\,h\,\text{Mpc}^{-1}$), which are of most interest for cosmological experiments since perturbations are more linear, the best performance comes from the interlaced PCS scheme as one would expect. On these scales, there is slight evidence for the hierarchical ordering of results. This is shown by \autoref{tab:AverageBias} where we show the average bias across these scales for each scheme. We see a near consistent gradual decrease in the bias for higher-order schemes with interlacing. We chose this scale range to avoid the largest scales where cosmic variance will start to dominate and potential masking/modelling effects may be less accurate, and also the smallest scales where the residual aliasing is having the most impact and would be discarded in real analysis. The \autoref{tab:AverageBias} results thus motivate the use of higher-order interpolations with interlaced fields, for high precision cosmology with LIM data. We also tested these processes on the Cubic voxels simulation where aliasing is more prevalent at lower-$k$ and found equally conclusive results concerning the need for damping and aliasing treatments.

\section{Discussion \& Conclusion}\label{sec:Conclusion}

Line intensity maps provide matter tracing fluctuations across a 3D space in \textit{sky} coordinates, where the intensity in each voxel corresponds to a position in (R.A., Dec., $\nu$). Analysing these fluctuations in Fourier space, e.g. with a 2-point power spectrum estimation, requires the sky map to be resampled onto a regularly spaced Cartesian grid to allow a computationally viable FFT to be performed. The voxel coordinates can be trivially transformed to Cartesian coordinates and their intensities binned into the new Cartesian cells. However, if the choice of resolution for the new Cartesian grid is too high, empty cells will appear in the field and create unwanted ringing effects in the Fourier transform. Too low a resolution artificially reduces the Nyquist frequency, limiting the scale range in which the analysis can be performed. 

In this work, we showed how a Monte-Carlo-style integration technique, whereby a number of sampling particles are assigned to each map voxel and then transformed to Cartesian space, is an effective way to resample LIM data. However, we showed how small-scale separations between the sampling particles can inject unwanted aliasing into a power spectrum estimation. Inspired by previous work \citep{Jing:2004fq,Cui:2008fi,Sefusatti:2015aex,Hand:2017pqn,Blake:2019ddd} which addresses a similar effect but mostly from gridding of point-like galaxies, we presented corrections that can be applied both at field-level and in the modelling to unbias the effects from the regridding to Cartesian space. We then explored higher-order mass assignment schemes and interlaced fields as techniques to suppress the aliasing introduced from the resampling particles. We found that these can be effective solutions and allow sub-percent accuracy relative to the modelled expectation. We found there was a less conclusive hierarchy of results for the schemes compared to other results from galaxy survey simulations \citep[e.g.][]{Jing:2004fq,Sefusatti:2015aex}, but attributed this to the more complex process LIM data goes through, containing an additional interpolation step in the map-making where the continuous data is discretised into sky voxels. The complex mix of this initial interpolation, followed by the further interpolation at the Cartesian regridding stage is responsible for a less efficient improvement in accuracy from higher-order schemes. However, our results from \autoref{fig:Pk_mas_comparison} and \autoref{tab:AverageBias} show that the PCS scheme with an interlaced field delivered the most accurate results, well within sub-percent agreement for the scales of interest.

Whilst the process of resampling to a Cartesian grid can be avoided by using a harmonic-space $C_\ell$ power spectrum or a configuration-space correlation function, these estimators come with additional challenges and are less utilised in current LIM analyses. For example, a Fourier-space power spectrum analysis is adopted in \citet{Masui:2012zc,Anderson:2017ert,Keenan:2021uue,Wolz:2021ofa,COMAP:2021sqw,Cunnington:2022uzo}. In all these works some transformation to Cartesian space is required. In many cases, this is simplified by assuming the survey footprint is approximately rectangular in Cartesian space and assigning Cartesian equal dimensions to all the sky voxels in the map. However, as we showed, this assumption becomes increasingly inaccurate for wider sky surveys, already causing ${>}\,20\%$ bias for our $20\,{\times}\,20\,\text{deg}^2$ simulated survey. Experiments like MeerKLASS \citep{MeerKLASS:2017vgf} are now gathering data for a ${>}\,4{,}000\,\text{deg}^2$ radio intensity mapping survey, thus currently adopted approximations will no longer be suitable. The highly adaptable pipeline we have presented in this work provides the alternative framework for accurate analysis with future LIM.

The techniques we have outlined are all compatible with the other data reduction techniques necessary in LIM analysis. For example, foreground cleaning is a major part of an unbiased LIM pipeline and one we have not discussed in this work. However, blind foreground removal techniques \citep{Alonso:2014dhk,Carucci:2020enz,Cunnington:2020njn,Spinelli:2021emp} can be trivially performed on the original sky map data where the continuum foreground remains coherent along the frequency (line-of-sight) direction, for optimal component separation. Once foreground cleaning is performed, the regridding steps we have outlined in this work can then be performed with no modification necessary. Furthermore, the current adopted technique of using a transfer function to correct for signal loss in the foreground clean \citep{Cunnington:2023jpq} can still be implemented. The only change needed here would involve emulating the same regridding step that was performed on the data, on the injected mocks used for the transfer function computation. Whilst we have not explicitly checked these assumptions, we see no obvious reason that our regridding and modelling techniques will cause unforeseen problems to LIM foreground cleaning. We mention this to motivate an explicit check with future work. 

The field-level corrections we presented will be directly applicable to higher-order $n$-point estimators such as the bispectrum \citep{Randrianjanahary:2023rgp}\textcolor{black}{, and also one-point statistics \citep{Bernal:2023ovz}}. Furthermore, much of the modelling of discretistaion effects can be tweaked slightly to apply to high-order estimators, as similarly shown in \citet{Cunnington:2021czb} for modelling observational effects in the bispectrum. \textcolor{black}{It would also be interesting to see if the higher-order assignment schemes deliver a clearer improvement in performance for these statistics which are more sensitive to non-linear and non-Gaussian fluctuations.}

In addition to validating the claim that foreground removal presents no additional challenges for the regridding process (or vice-versa), we intend to further expand our investigation in future work. This work used simple lognormal mocks cut to a survey footprint. For a more robust test of the pipeline, we aim to extend these simulations to include a redshift evolving lightcone \citep[as in][]{Sato-Polito:2022wiq} and to also assign luminosities following a well-motivated luminosity function \citep[as in][]{Niemeyer:2023yeu}. These more realistic simulations will also introduce non-linear effects, boosting power on small scales, thus potentially enhancing aliased power. Furthermore, as discussed in \appref{sec:BeamEffects}, anisotropic observational effects such as a non-Gaussian, frequency-dependent beam will present additional challenges for LIM analysis, not just for regridding and aliasing mitigation techniques. We have also not included contributions to the field from thermal noise, residual foregrounds or systematics. It will also be necessary to include the case where LIM data are cross-correlated with galaxy surveys. These surveys should exhibit different aliasing contributions and hence validating that we can simultaneously handle their effects will be essential for future precision cosmology which will benefit from LIM-galaxy cross-correlations. 

The \texttt{Python} code which performs the LIM simulations, regridding, higher-order interpolations, power spectra estimation and modelling discussed in this paper, is made available at \href{https://github.com/stevecunnington/gridimp}{\texttt{github.com/stevecunnington/gridimp}}.

\section*{Acknowledgements}

The authors would like to thank Chris Blake and Alkistis Pourtsidou for useful discussions in the development of this project. We are also grateful for feedback on the final draft from José Luis Bernal, Isabella Carucci and Keith Grainge. 
Furthermore, we appreciate helpful comments and questions from Matilde Barberi Squarotti, Phil Bull, Melis Irfan, Aishrila Mazumder, Sourabh Paul, and Mario Santos.

SC is supported by a UK Research and Innovation Future Leaders Fellowship grant [MR/V026437/1]. LW is a UK Research and
Innovation Future Leaders Fellow [MR/V026437/1]. We acknowledge the use of the ilifu cloud computing facility (\href{www.ilifu.ac.za}{ilifu.ac.za}), a partnership between the University of Cape Town, the University of the Western Cape, Stellenbosch University, Sol Plaatje University and the Cape Peninsula University of Technology. The ilifu facility is supported by contributions from the Inter-University Institute for Data Intensive Astronomy (IDIA – a partnership between the University of Cape Town, the University of Pretoria and the University of the Western Cape), the Computational Biology division at UCT and the Data Intensive Research Initiative of South Africa (DIRISA).

For the purpose of open access, the authors have applied a Creative Commons Attribution (CC BY) licence to any Author Accepted Manuscript version arising.

\section*{Data Availability}

The data underlying this article will be shared on reasonable request to the corresponding author.



\bibliographystyle{mnras}
\bibliography{Bib} 




\appendix

\section{Resampled normalisation}\label{sec:ResampledNormalisation}

To find the correct normalisation factor $A$ for the intensities assigned to the sampling particles we can consider a simple case of resampling from one Cartesian grid $\delta_0$ to the final one with fewer cells $\delta_\text{G}$. The \textit{total} summed intensity appearing in the regridded field, which collects $N_\text{p}$ sampling particles for all $n_\text{0}$ equal-volume cells in the input grid, $\delta_0$, is given by
\begin{equation}\label{eq:massregridfield}
    \sum_i\delta_\text{G}(x_i) = \frac{N_\text{p}\sum_i (\delta_0(x_i)/A)}{N_\text{p}n_0/n_\text{G}}
\end{equation}
where $n_\text{G}$ is the number of cells in the final grid on which the FFT will be performed. The numerator here is the total mass assigned to all sampling particles with the normalisation factor $A$ that we intend to solve for. The denominator represents the average count of particles in cells for the new grid, simply the total particles, $N_\text{p}n_0$, divided by the number of new cells $n_\text{G}$.

When resampling to a new grid, we aim to leave the power spectrum amplitude, which measures the normalised fluctuations in the density fields, unchanged i.e. we are aiming for $P_0\,{=}\,P_\text{G}$. Defining the power spectrum at a certain scale as $P(k)\,{=}\,V\sigma(k)^2$ we can demand
\begin{equation}
    V_0\sigma_0^2 = V_\text{G}\sigma_\text{G}^2\,,
\end{equation}
where $V$ represents the cell volume for the particular grid. Since the overall volume for the whole field will remain the same this requirement becomes
\begin{equation}
    \frac{\sigma_0^2}{n_0} = \frac{\sigma_\text{G}^2}{n_\text{G}}\,.
\end{equation}
Defining the variance of the fluctuations fields as $\sigma^2\,{=}\,\sum(\delta-\bar{\delta})^2/n$, and considering that our fields are mean-centred, hence $\bar{\delta}=0$, we can write
\begin{equation}
    \frac{\sum(\delta_0^2)/n_0}{n_0} = \frac{\sum(\delta_\text{G}^2)/n_\text{G}}{n_\text{G}}\,.
\end{equation}
Making use of \autoref{eq:massregridfield} we can say
\begin{equation}
    \frac{\sum(\delta_0^2)}{n_0^2} = \left(\frac{\sum_i (\delta_0(x_i)/A)}{n_0/n_\text{G}}\right)^2 \Bigg/ n_\text{G}^2\,.
\end{equation}
From this we can see that for any change in grid size, the correct normalisation for the resampling particles must be $A\,{=}\,1$. This is why the intensities assigned to the sampling particles are not re-scaled.

\section{Input cosmology \& power spectra}\label{sec:FidCosmo}

For our simulated data we first generate a linear matter power spectrum assuming a Planck18 cosmology \citep{Aghanim:2018eyx} for which we use the CLASS Boltzmann solver \citep{Lesgourgues:2011re,Blas:2011rf} via \texttt{classylss}\footnote{\href{https://classylss.readthedocs.io/en/stable/}{classylss.readthedocs.io/en/stable/}}. As discussed in this work we choose to emulate a typical \hi\ (21cm) intensity map field, but this choice gives no loss of generality to our conclusions and should be applicable to any LIM experiment. From the CLASS matter power spectrum $P_\text{m}$, we can generate lognormal mocks \citep{ColesLognormal1991} for the \hi\ fluctuation field, $\delta_{\hi}(\boldsymbol{x},z)\,{=}\,b_\hi(z)\delta_\text{m}(\boldsymbol{x},z)$, by sampling the \hi\ power spectrum,
\begin{equation}
    P_\hi(\boldsymbol{k},z) = b_\hi^2(z) P_\text{m}(\boldsymbol{k},z)\,.
\end{equation}
In this work we neglect the impact of redshift-space distortions (RSD). Including RSD would require generating a potential field and projecting peculiar velocities along the \textit{true} line-of-sight, i.e. in the frequency direction. We will pursue this upgrade to the simulations in future work but do not foresee it changing the conclusions regarding regridding or aliasing effects, only introducing an additional modelling requirement. For the \hi\ bias we extrapolate a model based on hydrodynamical simulations \citep{Villaescusa-Navarro:2018vsg};
\begin{equation}\label{eq:HIbias}
    b_\hi(z) = 0.842 + 0.693z - 0.046z^2\,.
\end{equation}
The raw observed field for LIM data is typically a temperature fluctuation map in units of mK,
\begin{equation}
    \delta T_\hi(\boldsymbol{x},z) = \overline{T}_\hi(z)\delta_\hi(\boldsymbol{x},z)\,.     
\end{equation}
The mean \hi\ temperature is related to \hi\ density abundance by \citep{Battye:2012tg}
\begin{equation}\label{eq:TbarModelEq}
    \overline{T}_\hi(z) = 180\Omega_{\hi}(z)h\frac{(1+z)^2}{H(z)/H_0} \, {\text{mK}} \, ,
\end{equation}
where we assume $\Omega_\hi\,{=}\,0.83\,{\times}\,10^{-3}$ which was the measurement in \citet{Cunnington:2022uzo} at $z\,{=}\,0.43$, similar to what we assume in this work.

\section{Instrumental beam effects}\label{sec:BeamEffects}

Robustly simulating the beam for the purposes of an investigation into discretisation and aliasing effects is a challenging task. The beam is typically emulated in simulated data by smoothing the field once it has been generated. This smoothing should be perpendicular to the line-of-sight thus is most naturally applied on the sky maps in \texttt{HEALPix} coordinates. The problem however, is that if we want to preserve the effects from discretisation into sky voxels, then smoothing the field after it has undergone such discretisation will massively supress the effects and not emulate a realistic situation. This is why we chose to smooth the original $\delta_0$ input field in Cartesian space, making the assumption that the x-y plane is perpendicular to the true line of sight along frequency, which will not be the case for a wide survey with curved sky effects. We propose two alternative ways to improve this investigation in future work;
\begin{enumerate}[wide, labelwidth=!, labelindent=0pt]
    \item Smooth the input $\delta_0$ Cartesian field with a curved sky kernel model for the beam.
    \item Transform $\delta_0$ input to a high \texttt{nside} \texttt{HEALPix} map. Apply the beam in this space. Then re-pixelise to courser resolution with \texttt{nside} that emulates the real analysis.
\end{enumerate}
Method (i) should perfectly emulate reality providing that a curved sky beam kernel can be formulated, which is no trivial task. Method (ii) is analytically simpler but would require more computational demand since the simulation pipeline will now require higher \texttt{nside} maps and the additional resampling to the final resolution. The idea behind the latter method is that the discretisation onto a high \texttt{nside} map will produce negligible discretisation so applying the beam here will not artificially suppress the effects we wish to investigate. Introducing an enhanced beam model beyond a Gaussian profile will also increase the realism of the simulation, as explored in previous work \citep[e.g.][]{Matshawule:2020fjz,Spinelli:2021emp}. We leave the pursuit of these challenges to future work.

\begin{figure}
    \centering
    \includegraphics[width=1\linewidth]{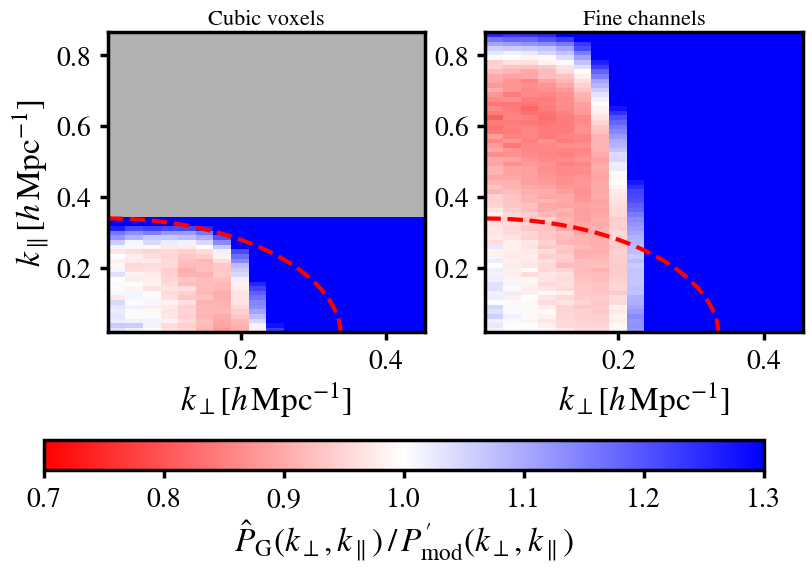}
    \caption{Cylindrical ($k_\perp, k_\parallel$) power spectrum comparison to model for two simulation versions. Same as \autoref{fig:Pk2D_alias} but with smoothing effects from the telecope beam. The red dashed line shows the \textit{minimum} Nyquist frequency calculated from the cell resolution along one of the dimensions in the perpendicular directions. Colour-bar range is fixed to avoid saturation.}
    \label{fig:Pk2D_alias_wbeam}
\end{figure}

\begin{figure}
    \centering
    \includegraphics[width=1\linewidth]{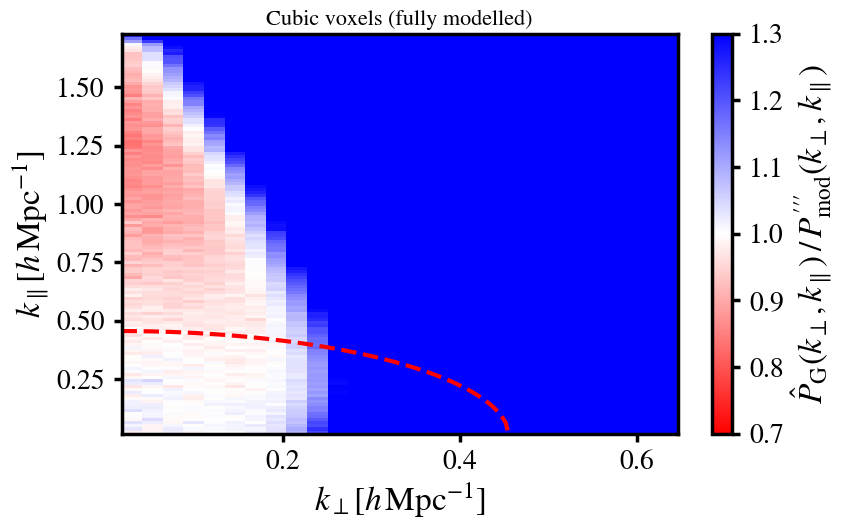}
    \caption{Cylindrical ($k_\perp, k_\parallel$) power spectrum for the Cubic voxels simulation (with the beam) relative to the final full model $P^{\prime\prime\prime}_\text{mod}$ (\autoref{eq:sumovermodes}) including the treatments for discretisation and aliasing from the sky map voxels. This shows the 2D version of \autoref{fig:Pk_mas_comparison} where PCS and interlacing has been used in the regridding of the data. The red dashed line shows the \textit{minimum} Nyquist frequency calculated from the cell resolution along one of the dimensions in the perpendicular directions. Colour-bar range is fixed to avoid saturation.}
    \label{fig:Pk2D_alias_fullmodel}
\end{figure}

With the caveat that our application of the beam is a somewhat simplified process, we found some interesting effects when analysing the distortions and aliasing in the anisotropic 2D $P(k_\perp,k_\parallel)$ power spectra. \autoref{fig:Pk2D_alias_wbeam} shows another version of \autoref{fig:Pk2D_alias}, comparing the model agreement between the Cubic voxels and Fine channels simulation, but now with the beam applied. We we see a similar conclusion that the discrepancies between measurement and model are pushed to higher $k_\parallel$ for the Fine channels simulation, beyond the lowest Nyquist frequency (red dashed line) determined from the perpendicular directions. However, at the higher $k_\parallel$ resolved by the grid, we now see lots of under-estimated power relative to the model (red regions) which we did not see in the resolved modes without the beam (\autoref{fig:Pk2D_alias}). The power spectrum modelling we use should include a perfect representation of the beam that we applied to the Cartesian field in the simulations, therefore it is difficult to explain the source of the negatively biased modes. The fact that this negative discrepancy is most pronounced for the high-$k_\parallel$ modes in the Fine channels version is also interesting since the beam is damping high-$k_\perp$ modes. Evidence was presented in the \citet{Villaescusa-Navarro:2016kbz} to show the damping effect from the beam can impact parallel scales. This was in the context of parallel baryon acoustic oscillations in the presence of a large beam, but it could be related to what we see. Further investigation with a more realistic beam is warranted in future work to fully understand this effect. For now, it is sufficient to see that when we include the full modelling treatments onto a more resolved Cartesian grid (see \autoref{fig:Pk2D_alias_fullmodel}, following steps in \secref{sec:MitigatingAliasingContributions}) we see excellent agreement for the area of $k$-space ($k\,{\lesssim}\,0.2\,h\,\text{Mpc}^{-1}$) of interest for the 21cm LIM survey we have simulated.

\bsp	
\label{lastpage}
\end{document}